\documentclass[a4paper,11pt]{article}
\pdfoutput=1 % if your are submitting a pdflatex (i.e. if you have
\usepackage{jinstpub} % for details on the use of the package, please
\usepackage{hyperref}
\usepackage{float}
\usepackage{graphicx}
\usepackage{subfigure}
\usepackage{leftidx}
\usepackage{multirow}
\usepackage{booktabs}
\usepackage{microtype}
\usepackage{threeparttable}  
\usepackage{amsmath}
\usepackage{amssymb}
\usepackage{enumerate}
\usepackage{cleveref}
\title{\boldmath A comparative study on different background estimation methods for extensive air shower arrays}
\author[a]{Yan-Jin Wang,}
\author[b,1]{Min Zha,\note{Corresponding author.}}
\author[b,d]{Shi-Cong Hu,}
\author[c]{Chuan-Dong Gao,}
\author[e]{Jian-Li Zhang,}
\author[a,f,g,1]{Xin Zhang}

\affiliation[a]{College of Sciences, Northeastern University, \\Shenyang 110819, China}
\affiliation[b]{Key Laboratory of Particle Astrophysics, Institute of High Energy Physics, Chinese Academy of Sciences, \\Beijing 100049, China}
\affiliation[c]{Institute of Frontier and Interdisciplinary Science, Shandong University, \\Qingdao 266237, China}
\affiliation[d]{University of Chinese Academy of Sciences, \\Beijing 10049, China}
\affiliation[e]{National Astronomical Observatories, Chinese Academy of Sciences, \\Beijing 100101, China}
\affiliation[f]{Key Laboratory of Data Analytics and Optimization for Smart Industry (Northeastern University), Ministry of Education, \\Shenyang 110819, China}
\affiliation[g]{National Frontiers Science Center for Industrial Intelligence and Systems Optimization, Northeastern University, \\Shenyang 110819, China}

\emailAdd{zham@ihep.ac.cn}
\emailAdd{zhangxin@mail.neu.edu.cn}

\abstract{Background estimation is essential when studying TeV $\gamma$-ray astronomy for extensive air shower arrays. In this work, by applying four different methods including equi-zenith angle method, surrounding window method, direct integration method, and time-swapping method, the number of the background events is calculated. Based on simulation samples, the statistical significance of the excess signal from different background estimation methods is determined. Following this, we discuss the limits and the applicability of the four methods under different conditions. Under the detector stability assumption with signals, the results from the above four methods are consistent at the 1$\sigma$ level. In the no signal condition, when the acceptance of the detector changes with both space and time, the surrounding window method is most stable and hardly affected. In this acceptance assumption, we find that the background estimation in the direct integration and time-swapping methods are sensitive to the selection of time window, and the shorter time window can reduce the impact on the background estimation to some extent.}
\keywords{background estimation method, detector stability, acceptance variation function}

\begin{document}
\maketitle
\flushbottom

%%%%%%%%%%%%%%%%%%%%%%%%%%%%%%%%%%%%%%%%%%%%%%
%%%%%%%%%%%%%%%%%%%%%%%%%%%%%%%%%%%%%%%%%%%%%%
%%                                          %%
%%  main contents                           %%
%%                                          %%
%%%%%%%%%%%%%%%%%%%%%%%%%%%%%%%%%%%%%%%%%%%%%%
%%%%%%%%%%%%%%%%%%%%%%%%%%%%%%%%%%%%%%%%%%%%%%
%=================================================================
\section{Introduction}\label{sec-1}
In 1989, the first successfully monitored signal from the Crab Nebula by Imaging Air Cherenkov Telescope (IACT)~\cite{Weekes:1989tc}   has opened a window into TeV $\gamma$-ray astronomy. Complementing the IACT, the Extensive Air Shower (EAS) experiment with the advantage of all day operation, weather independence, and a wide field of view providing simultaneous monitoring all sources within its field of view becomes a very important technique, especially in the search for transient phenomenon. 

Typically for the EAS experiment, a standard method to search for a signal from a point source in a certain sky-region is analyzed by comparing the number of events from the direction with the expected number of background events, then a statistical formula is used to calculate the possible excess~\cite{Alexandreas:1992ek,Li:1983fv}. Usually the observed $\gamma$-ray flux is very weak, around 0.1\%, compared with the background events from cosmic ray hadronic showers. Thus, in search of the signal, background estimation is very important to insure that any observed excess is not a background fluctuation or systematic effect introduced by the detector or analysis method.
 
The purpose of this paper is to discuss the uncertainty introduced by background estimation methods. The limits of their applicability on different conditions are also discussed. Based on Monte Carlo (MC) samples, the dependence of the methods on detector stability are also presented. The paper is organized as follows. The steps involved in the detailed simulation conditions will be described in section~\ref{sec-2}. Four different background estimation methods and their detailed implementation conditions will be introduced in section~\ref{sec-3}. The detailed discussion due to different background estimation in the condition of detector stable acceptance will be presented in section~\ref{sec-4}. The various conditions relative to unstable acceptance will be given in section~\ref{sec-5}, followed by a short summary and discussion.

%=================================================================
\section{Simulations}\label{sec-2}
We simulate the cosmic ray background and $\gamma$ signal in the detector stability assumption that the acceptance of the detector does not vary with time and direction.
The simulation samples are generated according to the following procedure:

\begin{enumerate}
\item The number of recorded cosmic ray events is sampled with the rate of 1k Hz which is approximately the experimental event rate at its typical energy  (around 2 TeV)~\citep{LHAASO:2021ozi}. The number of events in unit time follows Poisson distribution.
\begin{figure}[H]
  \centering\includegraphics[width=0.49\linewidth]{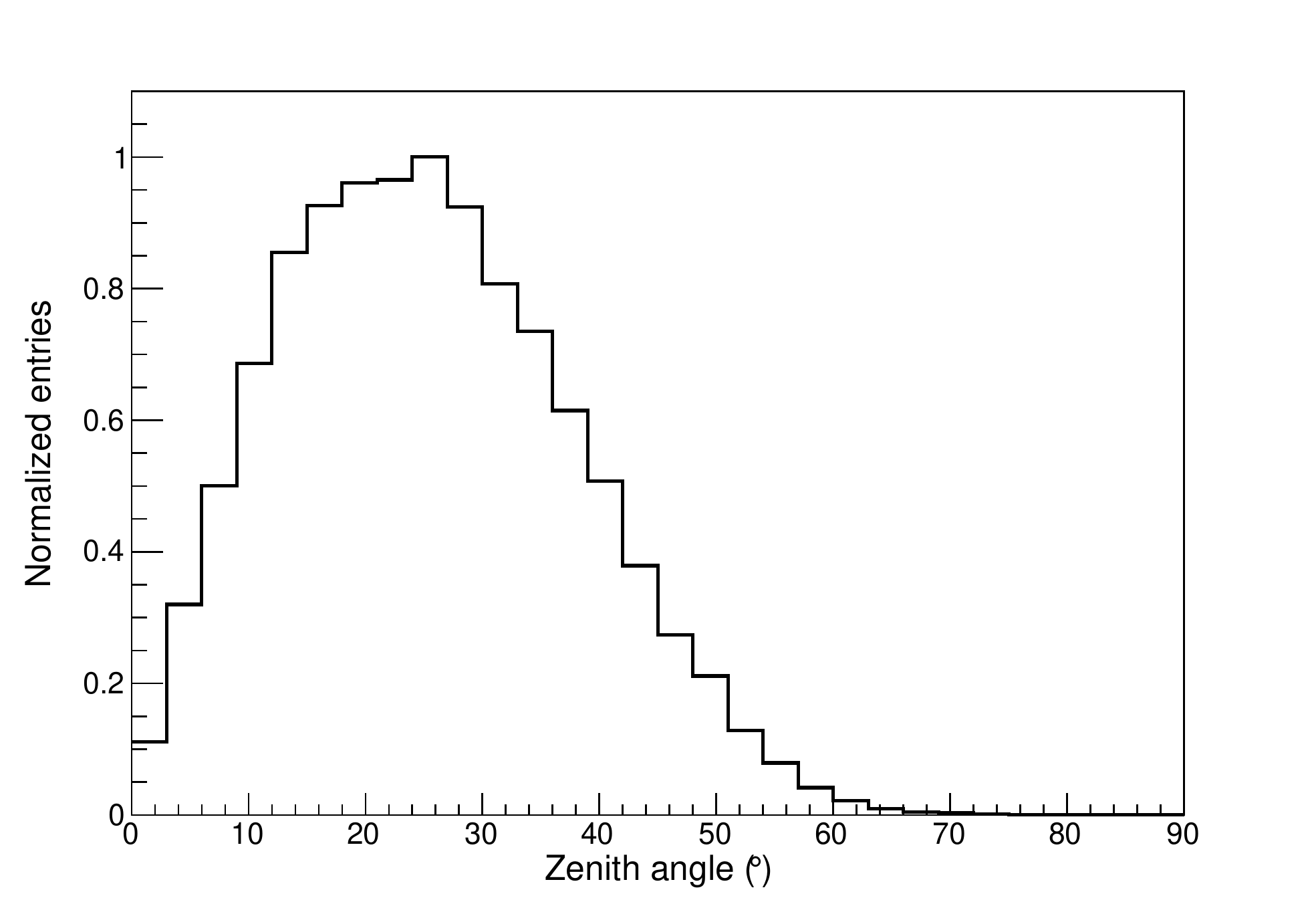}
  \centering\includegraphics[width=0.49\linewidth]{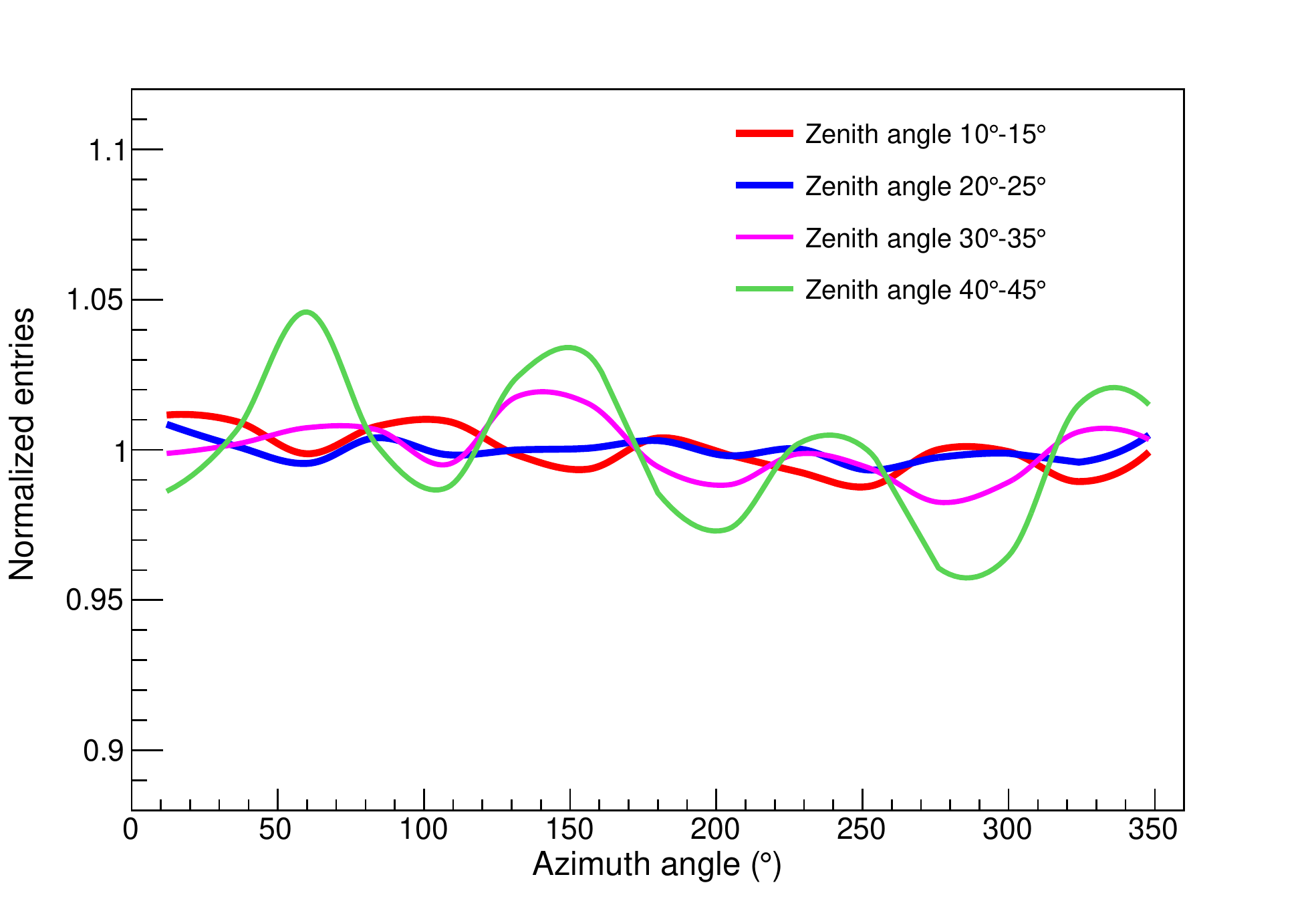}
  \caption{The distributions of zenith angle and azimuth angle from the simulation sample.
 Left panel: the zenith angle distribution from $0^{\circ}-90^{\circ}$.
 Right panel: the azimuth angle distribution in different zenith angle range. Red line is for the zenith angle in the range of $10^{\circ}-15^{\circ}$, blue line is for the zenith angle in the range of $20^{\circ}-25^{\circ}$, violet line is for the zenith angle in the range of $30^{\circ}-35^{\circ}$, and green line is for the zenith angle in the range of $40^{\circ}-45^{\circ}$. }
  \label{fig:zen_azi_distribution}
\end{figure}
\item The air shower arrival direction is directly sampled from the distribution given in the paper~\citep{LHAASO:2021ozi}. The detailed distribution of zenith angle is shown in the left panel of Figure~\ref{fig:zen_azi_distribution}. Events with a zenith angle less than $50^{\circ}$ are used in our analysis. The distributions of azimuth angle are divided into 10 groups in the zenith angle of $0^{\circ}$ to $50^{\circ}$, with a gap of $5^{\circ}$ in each group. 4 groups of the distributions are shown in the right panel of Figure~\ref{fig:zen_azi_distribution}. One can clearly see that the non-uniformity is less than $\pm 1\%$ within $10^{\circ}-15^{\circ}$, and 5\% as the zenith angle increases to the $40^{\circ}-45^{\circ}$ region. Besides temperature and barometric pressure, many other factors can cause the non-uniformity of azimuth, such as non-perfect symmetric detector shape, data quality cuts and so on.
\item A fake $\gamma$ signal with the Crab nebula orbit is generated. The intensity is tuned with the expected signal significance of $15\sigma$, which is close to the daily significance from the experimental observation. According to the experimental angular resolution result of LHAASO-WCDA in the paper~\citep{LHAASO:2021ozi}, in this simulation a $0.25^{\circ}$ angular resolution is added to each $\gamma$ event using the Point Spread Function (PSF).
\item All simulated events are spread uniformly during the time. 
\end{enumerate}
\begin{figure}[H]
  \centering\includegraphics[width=0.49\linewidth]{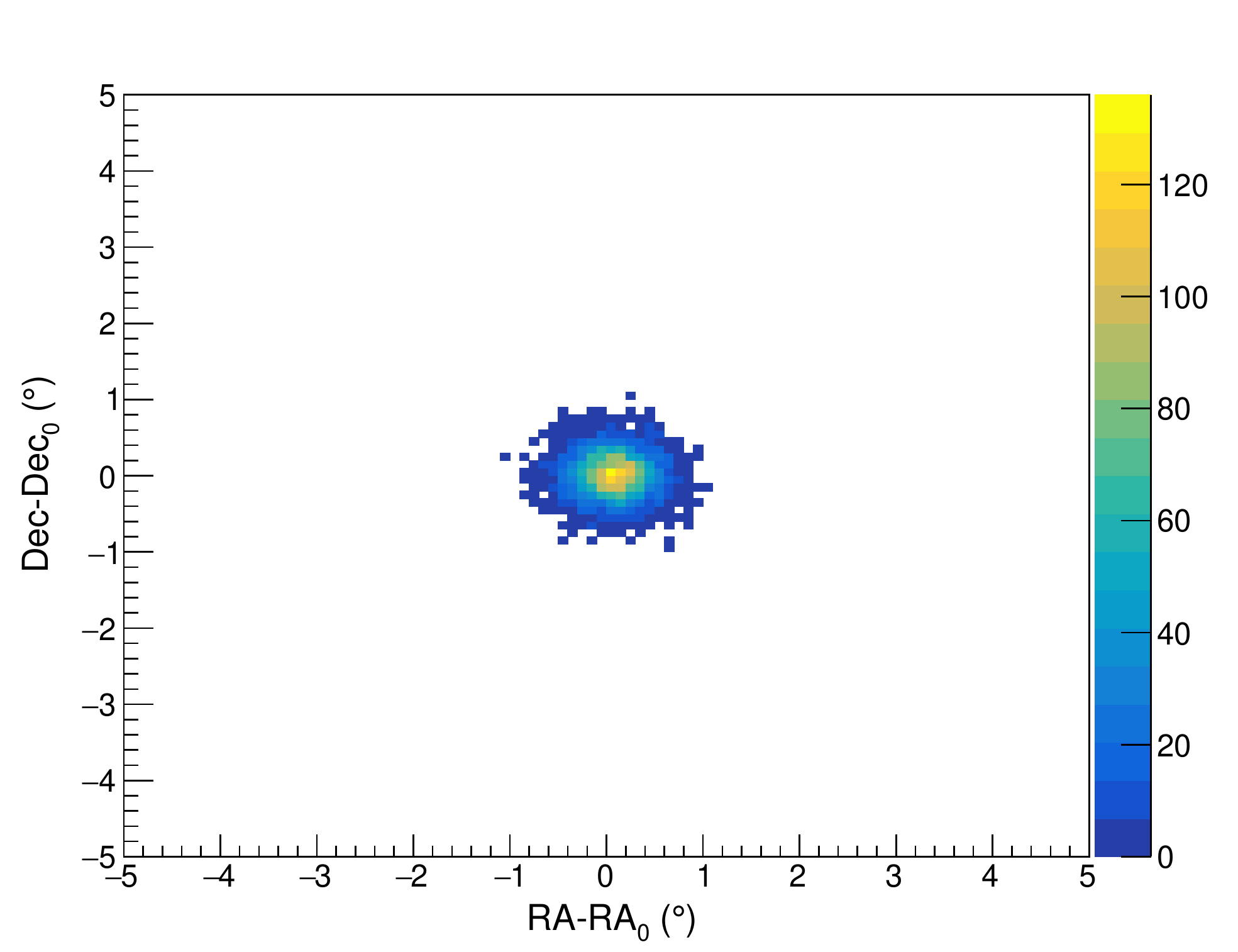}
  \centering\includegraphics[width=0.49\linewidth]{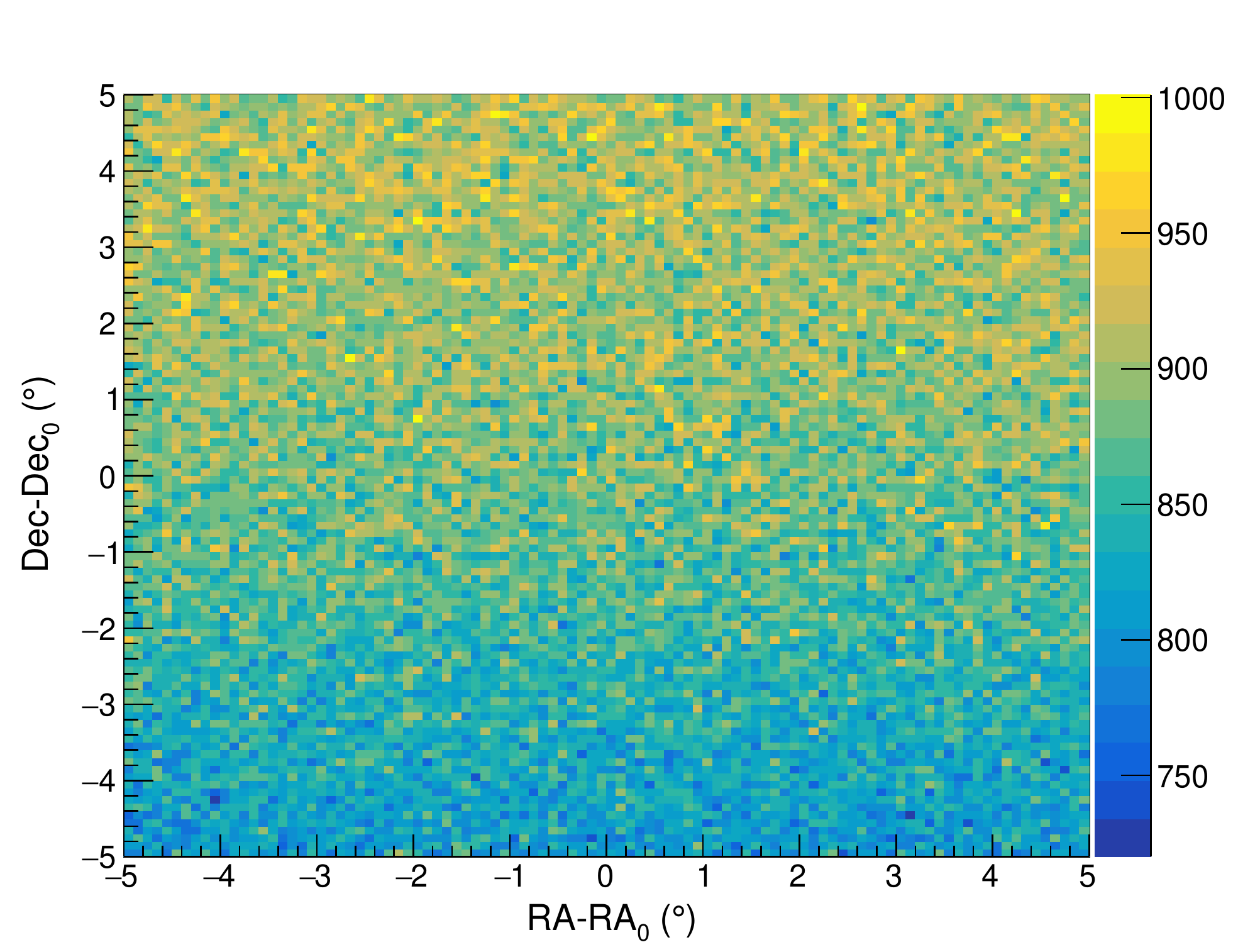}
  \caption{The generated sky maps from the region of the Crab. The left panel is the $\gamma$ signal map, with x and y coordinates representing the positions in right ascension (RA) and declination (Dec) directions relative to the Crab ($\rm{RA}_{0},\rm{Dec}_{0}$). The right panel is the cosmic ray background map, and the color scale represents the number of events in the pixel.}
  \label{fig:sig_back_distribution}
\end{figure}

By following the procedure described above, about $5.16\times10^{3}$ and $1.73\times10^{9}$ showers have been simulated for the $\gamma$ signals and cosmic rays, respectively. 
The expected sky maps from fake $\gamma$ signal and cosmic rays are presented in Figure~\ref{fig:sig_back_distribution}. The left panel is the pure $\gamma$ signal contribution, and the right panel is the map from cosmic rays. 
%=================================================================
\section{Background estimation method}\label{sec-3} 
\subsection{Coordinate systems}
\begin{figure}
  \centering\includegraphics[width=0.7\linewidth]{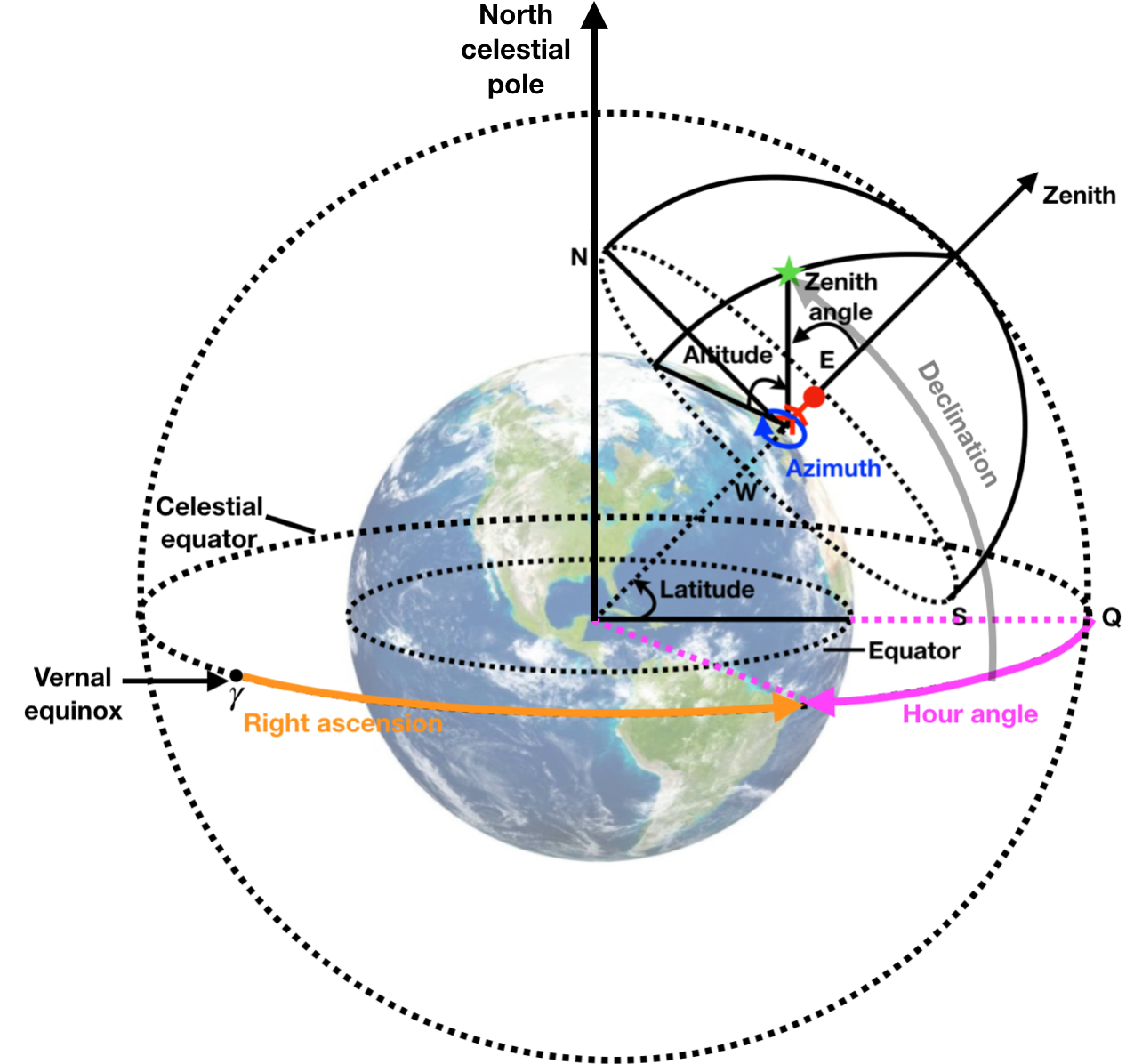}
  \caption{Celestial coordinate systems with zenith angle ($\theta$), azimuth angle ($\phi$), hour angle (HA), right ascension (RA), and declination (Dec).}
\label{fig:coordinates}
\end{figure}
In this paper, several coordinate systems are used to discuss the background estimation, such as horizontal coordinate system, hour angle coordinate system, and equatorial coordinate system, Figure~\ref{fig:coordinates} is a sketch map of these coordinate systems.  
\begin{enumerate}
\item The horizontal coordinate system is based upon the horizon of the observer, and it is made up of zenith angle ($\theta$) and azimuth angle ($\phi$).
$\theta$ is measured from $0^{\circ}$ to $90^{\circ}$ equivalently to the horizontal colatitude. 
$\phi$ is corresponding to the horizontal longitude from $0^{\circ}$ to $360^{\circ}$, which is measured from north towards east.
It is useful to know the position of the star relative to the observer, but it depends on the location of observation. 

\item The hour angle coordinate system is related to the observer, which is composed of hour angle (HA) and declination (Dec).
The origin of the HA is the point Q from Figure~\ref{fig:coordinates}, which is the intersection point of the celestial equator and meridian circle near the south point, and it is measured westwards in hours from -12 h to 12 h. In hour angle coordinate system, Dec is corresponding to the latitude, and its origin is the celestial equator from $-90^{\circ}$ to $90^{\circ}$.
This coordinate system depends on the location and time of observation, but the Dec of the star does not change.

\item The equatorial coordinate system consists of right ascension (RA) and Dec. The origin of RA is the vernal equinox $\gamma$ as shown in Figure~\ref{fig:coordinates}, and it is measured from $0^{\circ}$ to $360^{\circ}$ eastward. The relationship between RA and HA is given by $\rm{HA}=\rm{LST}-\rm{RA}$, where LST is the sidereal time. Dec is the same as that in the hour angle coordinate system. Different from the former two local coordinate systems, the equatorial coordinate system is independent of location and time of observation.
\end{enumerate}

The number of events expected inside a fixed region ($\theta,\phi$) over a period of time $t$ can be calculated by a function of $\theta$, $\phi$, and $t$, as $N_{\rm{b}}=\iiiint S \cdot F\cdot E^{-\gamma} \mathrm{d}E \mathrm{d}\cos(\theta)\mathrm{d}\phi dt$, where  $S$ is the effective area, $F$ is the flux, $E$ is the energy, and $\gamma$ is the energy spectral index. It shows that a very strong zenith angle dependence exists naturally in the data. Based on this fact, two different ways have been developed for the background estimation: a sky cell is used at the same time as the signal cell but in a different location; a second approach is to use a sky cell with same position as the signal cell but at a different time.

Below we describe in detail the four background estimation methods used in the EAS technique: equi-zenith angle method~\cite{Amenomori:2005pn}, surrounding window method~\cite{Adriani:2010zza}, direct integration method~\cite{Abdo:2011za}, and time-swapping method~\cite{Fleysher:2003nh}.

\subsection{Equi-zenith angle method}
In order to reduce the strong zenith angle dependence of the detector acceptance, the equi-zenith angle method is used to estimate background. 
The background is estimated by simultaneously counting the the number of events in off-source windows with different azimuth angle and in the same zenith as the on-source window. A sketch map of equi-zenith angle method is shown in Figure~\ref{fig:Equal_zen}. This method is not affected by the differences in atmospheric depth and it can obtain equal acceptance at equal zenith angle bands. In addition, the off-source windows in this method will not be contaminated by signals, which can make the background estimation more accurate. In this work, we take data within a certain time period covering $10^{\circ}\times 10^{\circ}$ around the source. We keep the ratio of on-source window to off-source window constant, here is 1/6. The expected number of background events in equi-zenith angle method can be expressed as
\begin{equation}
N_{\rm{b}}=\frac{\sum_{i=1}^{6}N_{\mathrm{off},i}}{6},
\end{equation}
where $N_{\mathrm{off},i}$ is the number of events in $i$th off-source window. In this method, two key factors should be noted:
\begin{figure}[H]
  \centering\includegraphics[width=0.46\linewidth]{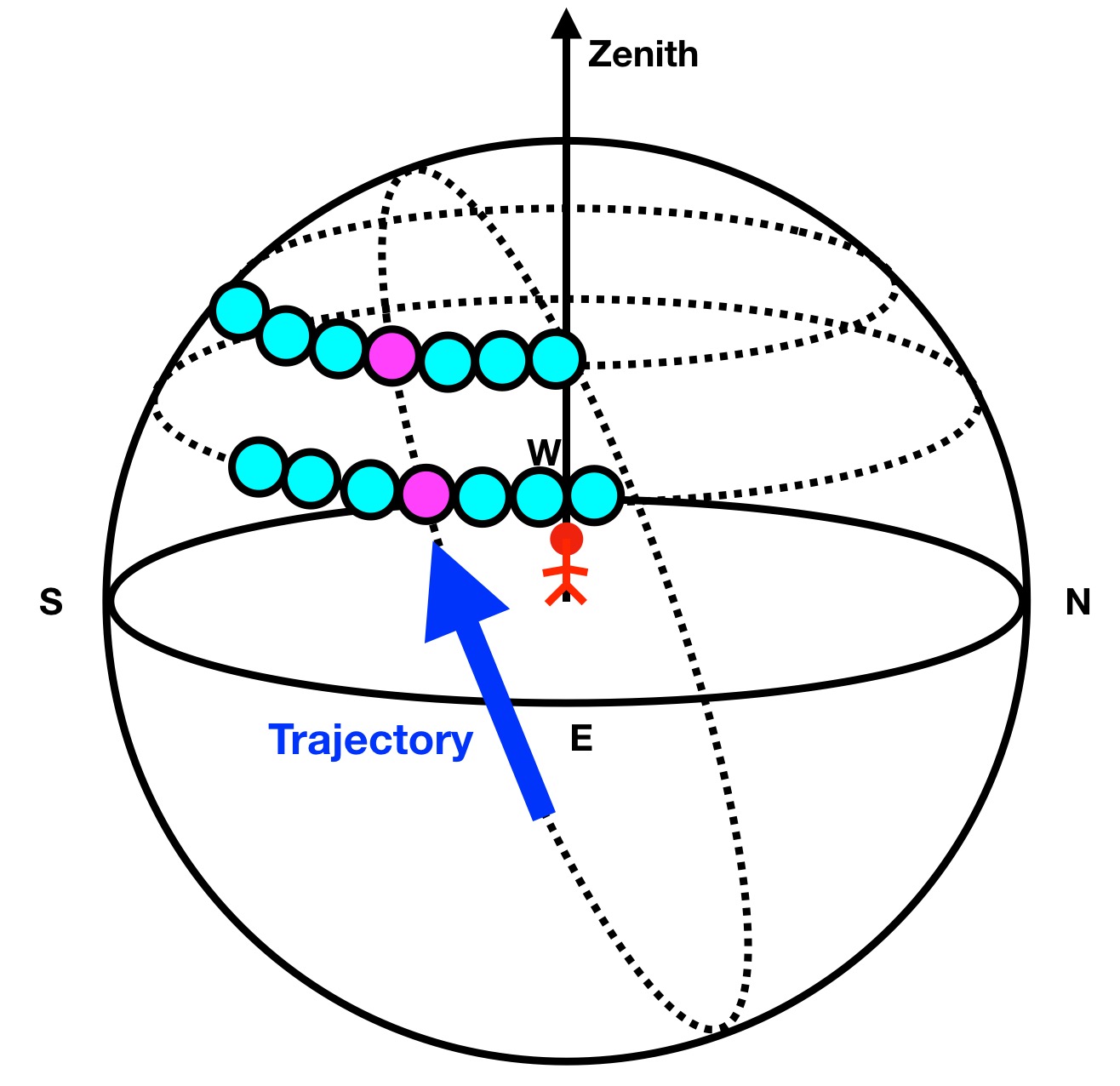}
  \centering\includegraphics[width=0.46\linewidth]{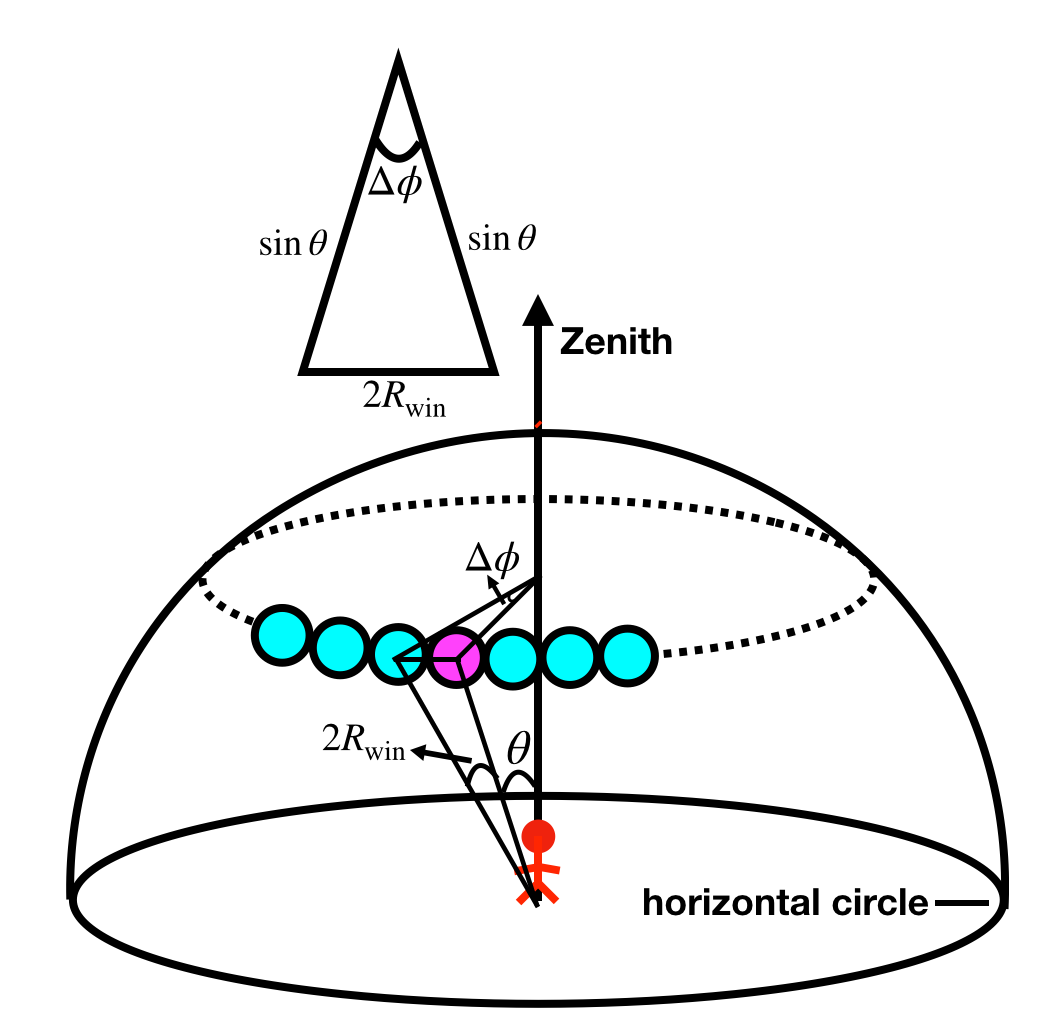}
  \caption{Sketch maps of the equi-zenith angle method. Left panel: the trajectory of the source, the on-source and off-source windows. In the same zenith angle band, off-source windows are symmetrically placed on both side of the on-source window. Pink region is the on-source window and light blue regions are off-source windows. The blue arrow is the trajectory direction of the source. Right panel: the relationship between zenith angle $\theta$, the angular radius of the on-source or off-source window(s) $R_{\rm{win}}$, and the azimuth angular distance of adjacent windows $\Delta \phi$.}
\label{fig:Equal_zen}
\end{figure}

\begin{enumerate}
\item To avoid the off-source window overlapping near small zenith angle, the limit condition meets:
\begin{equation}\label{dphi}
\Delta\phi(\rm{m}+1)\leq2\pi,
\end{equation}
where $\Delta \phi$ is the azimuth angular distance of adjacent windows, and m is the number of off-source windows. From the right panel of Figure~\ref{fig:Equal_zen}, we can obtain the relationship between $\theta$ and $\Delta \phi$ as follow:
\begin{equation}\label{theta_dphi}
\frac{R_{\rm{win}}}{\sin\theta}=\sin(\frac{\Delta\phi}{2}), 
\end{equation}
where $R_{\rm{win}} = 5^{\circ}$ is the angular radius of the on-source or off-source window(s). From equations~(\ref{dphi}) and~(\ref{theta_dphi}), we obtain a cut on zenith angle:
\begin{equation}\label{zencut} 
\sin\theta\geq\frac{R_{\rm win}}{\sin(\frac{\pi}{\rm{m}+1})}.
\end{equation}
According to equation~(\ref{zencut}), we can deduce that 6 independent $5^{\circ}$ off-source windows will require $\theta\geq11.6^{\circ}$, which also means that 14\% signal events are lost around small zenith.

\item The non-uniform event distribution of azimuth angle must be properly corrected during the analysis. A flat distribution of azimuth angle can be achieved by weighting events with normalized distribution of azimuth angle. More details about this method and correction can be found in~\cite{Amenomori:2005pn}. 
\end{enumerate}

In order to cover the full sky, all events in the equi-zenith angle belt except those inside the on-source window are taken as off-source events, and the ratio of on-source window to off-source window is different in different zenith angle belts. More details can be found in~\cite{Amenomori:2005pn}. 

\subsection{Surrounding window method}
\begin{figure}[H]
  \centering\includegraphics[width=0.8\linewidth]{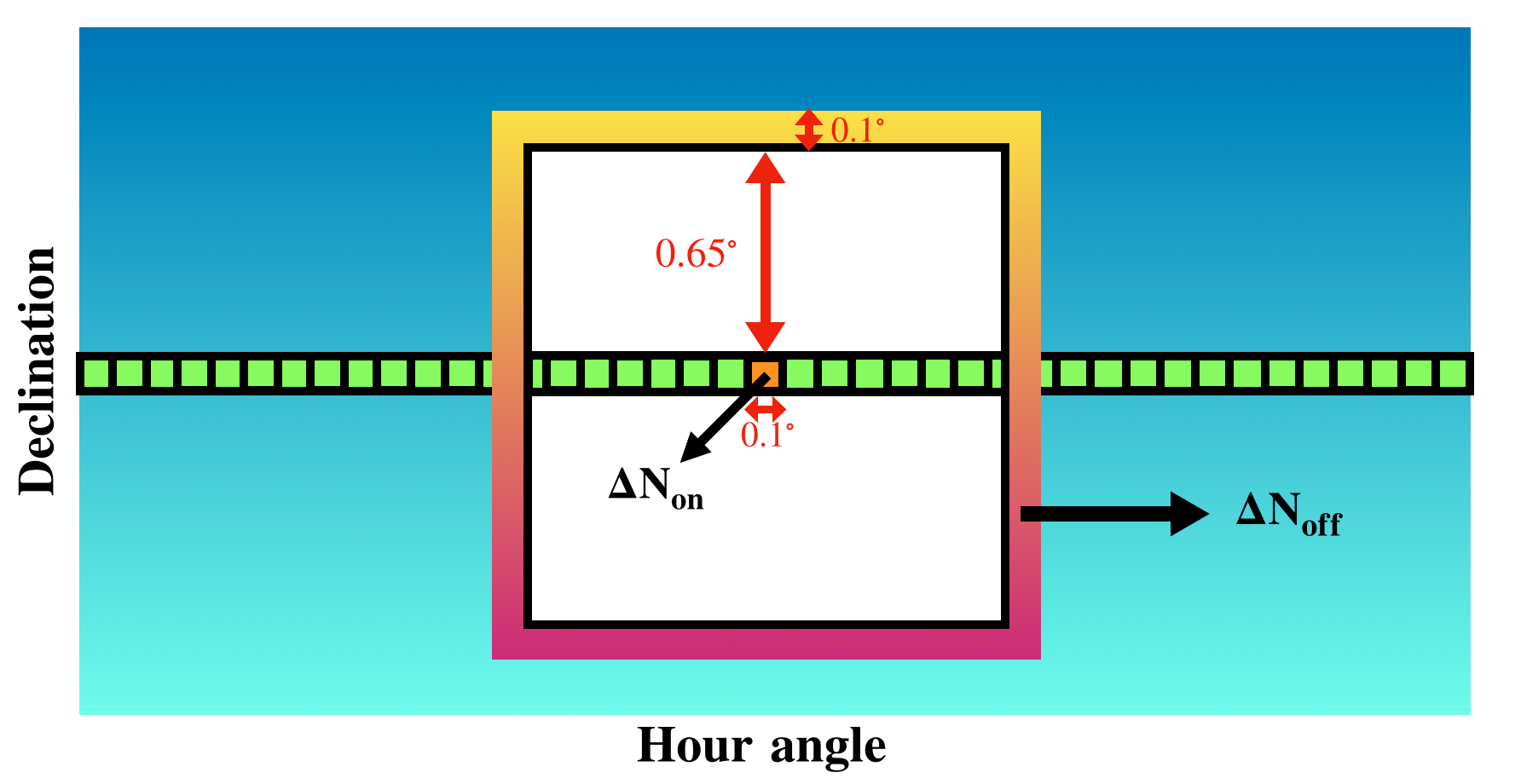}
  \caption{A sketch map of the surrounding window method. In hour angle coordinate system, a hollow rectangle around the sky cell is selected as the background region, and every sky cell has a corresponding background region surrounding it. One can estimate the background events in each cell from these hollow rectangles.}
  \label{fig:Surrounding_window_diagram}
\end{figure}
\begin{figure}
  \centering\includegraphics[width=0.58\linewidth]{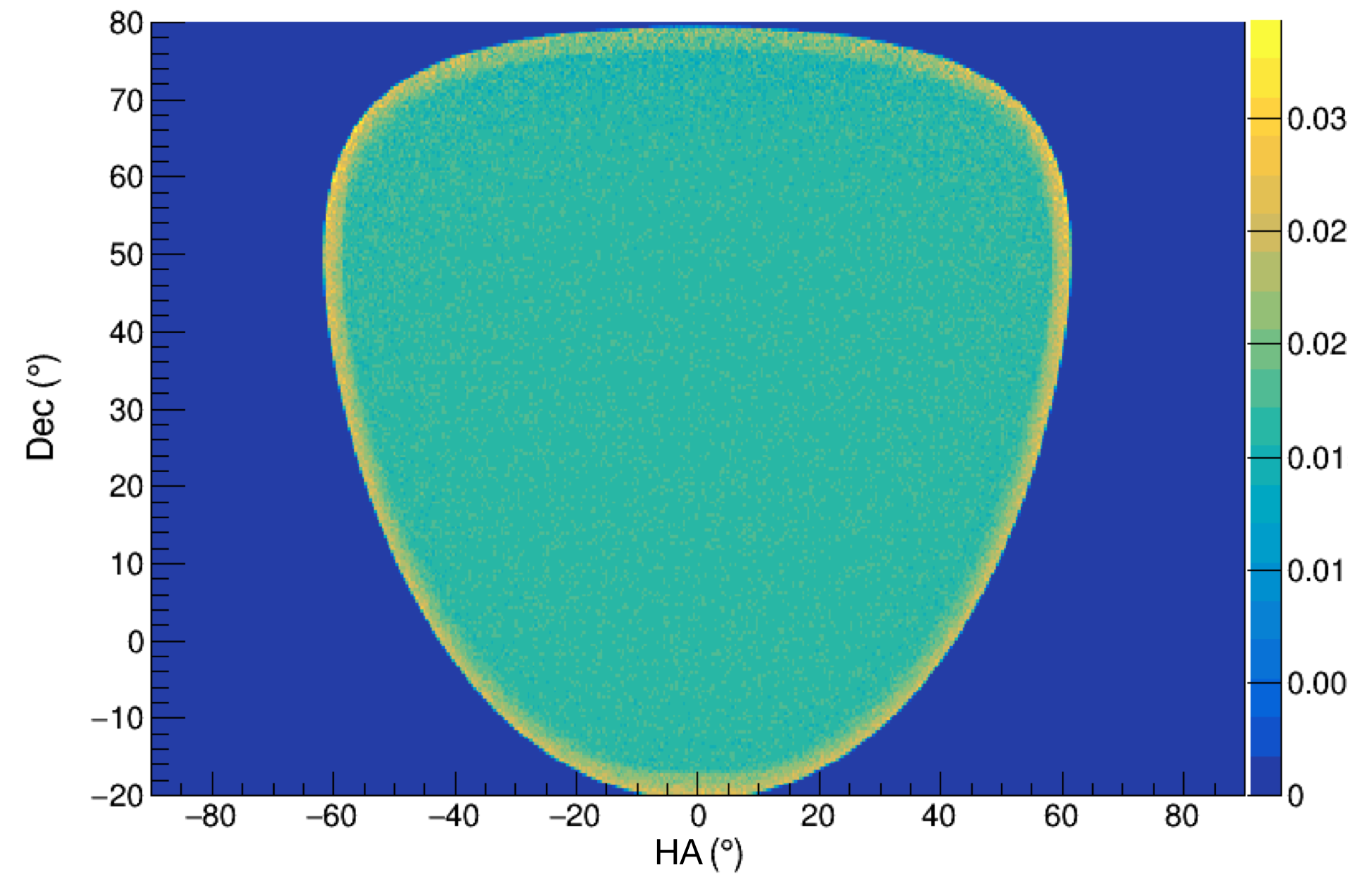}
  \caption{ The acceptance ratio for a 24 hour sidereal day in hour angle coordinates.}
  \label{fig:surrounding-window_acc_ratio}
\end{figure}

In the surrounding window method, a rectangular cell around the source is defined as signal region and a hollow rectangular region around the cell of interest is defined as the background region. A sketch map is shown in Figure~\ref{fig:Surrounding_window_diagram}. 
The principle of the method relies on the fact that the ratio, $R$, of events from a signal region to background region remains constant for a fixed direction with respect to the detector and can be measured accurately over the long period of the experiment. By counting the number of events in the surrounding region within the same time interval, and knowing $R$, the expected number of background events in the cell of interest can be calculated. Figure~\ref{fig:surrounding-window_acc_ratio} shows the acceptance ratio $R$ for the 24 hours of a sidereal day.

In this work, the width of the rectangle cell in declination is equally set as $\Delta \rm{Dec}=0.1^{\circ}$ and the relative width in HA is $\Delta \rm{HA}=\Delta \rm{Dec}$. For the background hollow rectangle, the half width in declination direction is from $\omega_{\rm{a}}=7\times\Delta \rm{Dec}$ to $\omega_{\rm{b}}=8\times\Delta\rm{Dec}$ and the half width in \rm{HA} direction is from $\omega_{\rm{a}}\times(\Delta\rm{HA}/\Delta\rm{Dec})$ to $\omega_{\rm{b}}\times(\Delta\rm{HA}/\Delta\rm{Dec})$. The number of background events for the surrounding window method in equatorial coordinates HA, Dec can be calculated by 
\begin{equation}
N_{\rm{b}}(\rm{RA},\rm{Dec})=R(\rm{HA},\rm{Dec})N_{\rm{off}}, 
\end{equation}
where $R(\rm{HA},\rm{Dec})$ is the acceptance ratio from a signal region to background region in HA and Dec, and $N_{\rm{off}}$ is the number of events in the background hollow rectangle.

\subsection{Direct integration method}
An assumption is needed for this method: the acceptance of the detector, $R^{\prime}(\theta, \phi, t)$, can be factorized as $R^{\prime}(\theta, \phi, t)= A(\theta, \phi)R(t)$. That is to say, firstly, the expected probability distribution of the background events in local coordinates is proportional to the event distribution observed in local coordinates $A(\theta, \phi)$; secondly, the background distribution remains nearly constant in a certain time duration. In this way, the expected number of background events can be estimated numerically by discretizing  $A(\theta, \phi)$ and $R(t)$ on a fine grid and replacing the integrations by sums.
\begin{figure}
  \centering\includegraphics[width=0.8\linewidth]{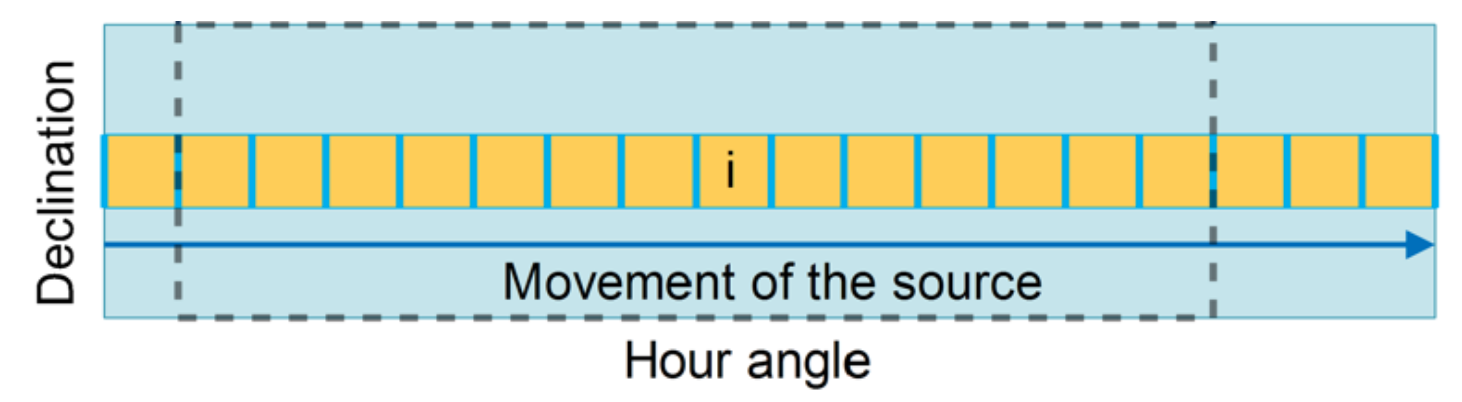}
  \caption{ A sketch map of the direct integration method. In hour angle coordinates, the source traverses an equal declination path. The sky is divided into small grids. If the source passes through the ith cell, the cell i is the signal region. If the source traverses the declination path earlier or later than the ith cell, the cell i is called background region. During this time, there is no signal in the cell i, which is used to estimate the background.}
\label{fig:direct_integral_diagram}
\end{figure} 

A sketch map of the direct integration method is shown in Figure~\ref{fig:direct_integral_diagram}. 
In hour angle coordinates, the sky map is divided into small grids and the number of events in each small time interval is recorded. 
Then the number of background events can be calculated by counting the number of events that fall on the grids in the same declination band.
Over a long observation time, the number of background events in equatorial coordinates RA, Dec can be calculated by numerical integration:
\begin{equation}\label{direct}
N_{\rm b}(\rm{RA},\rm{Dec}) = \int\int A( \mathrm{HA}, \rm{Dec})R(t)\epsilon\mathrm{d}\Omega \mathrm{d}t, 
\end{equation}
where $A(\rm{HA},\rm{Dec})$ is the acceptance distribution of the events in hour angle coordinates ($\rm{HA},\rm{Dec}$). $R(t)$ is the event rate of the detector as a function of time. 
$\epsilon$ is a factor to judge whether the background event is valid or invalid, and if a background event fall within the  $(\rm{RA},\rm{Dec})$ bin in sidereal time $t$, $\epsilon = 1$, otherwise $\epsilon = 0$. 

\begin{figure}
  \centering\includegraphics[width=0.6\linewidth]{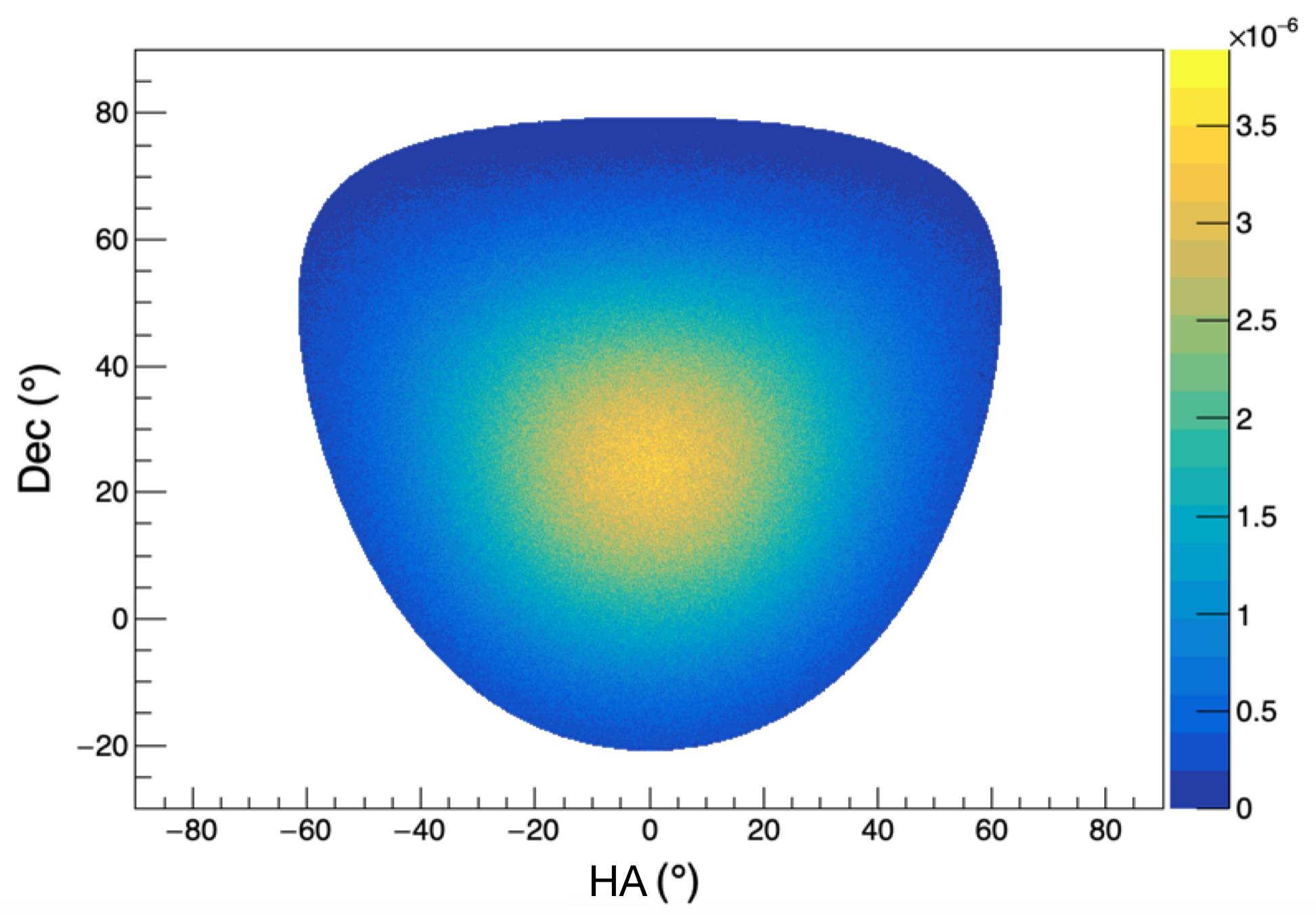}
  \caption{The acceptance integrated for a 24 hour sidereal day in local coordinates of hour angle and declination, the color is the normalized acceptance value.}
  \label{fig:fig5}
\end{figure}

In this analysis, the integration duration is no more than the 24 hours of a sidereal day. 
The acceptance distribution, $A(\rm{HA},\rm{Dec})$, can be established by counting the number of events which fall within each bin over 24 hours and is shown in Figure~\ref{fig:fig5}. It should be noted that for the direct integration method and time-swapping method (discussed later in the text), to avoid signal contamination, a $1^{\circ}$ region around the source region is excluded during the background estimation. 

\subsection{Time-swapping method}
\begin{figure}[H]
  \centering\includegraphics[width=0.58\linewidth]{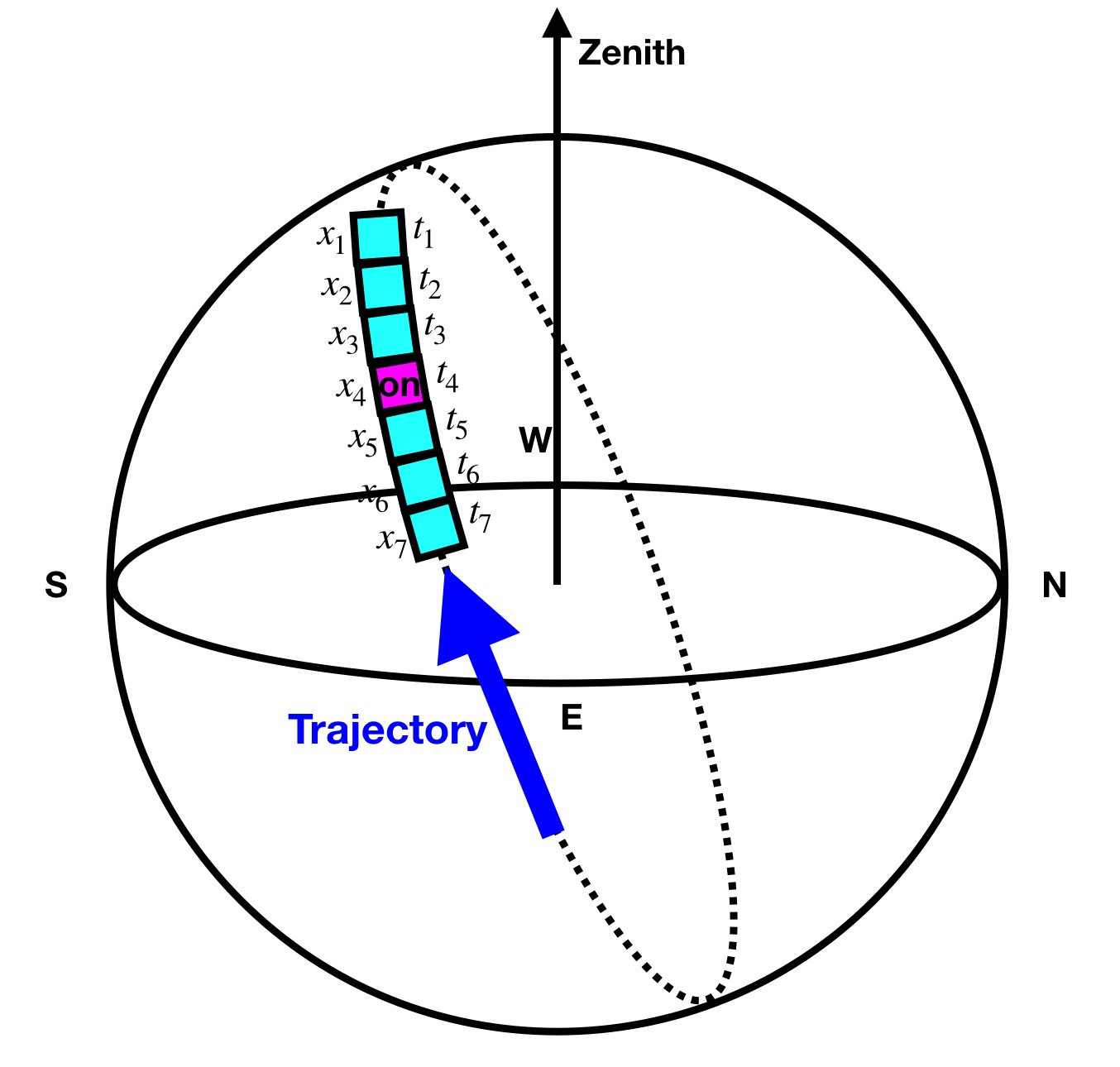}
  \caption{A sketch map of the time-swapping method. The pink region is source bin and the blue arrow is the trajectory direction of the source. The light blue regions are background bins estimated with the time-swapping method. There are different times in the background bins, and this method exchanges the time of background events at a given spatial location.}
\label{fig:Equal_dec}
\end{figure}

The principle of the time-swapping method is similar to the direct integration method, the difference is in its implementation. 
A sketch map of the time-swapping method is shown in Figure~\ref{fig:Equal_dec}. In hour angle coordinate system, the fake background events are created by the randomly combining the arrival time $t$ and the direction $x$ of the events within time window $T_{\rm w}$. 
This procedure can be repeated $N_{\rm{swap}}$ times. In this manner, the number of background events can be expressed as
\begin{equation}
N_{\rm b}= \frac{1}{N_\mathrm{swap}} \sum_{i=1}^{N}\epsilon(x_{i},t_{i}),
\end{equation}
where $N$ is the generated total number of fake events, and $\epsilon(x_{i},t_{i})$ is a factor to judge whether the fake background event within $T_{\rm{w}}$ is valid ($\epsilon(x_{i},t_{i}) = 1$) or invalid ($\epsilon(x_{i},t_{i}) = 0$). The larger $T_{\rm{w}}$, the better statistics can be accumulated and the more accurate acceptance distribution of the events can be achieved. However, with the increasing time window, the assumption of time independence of event acceptance becomes less valid. Here $T_{\rm{w}}$ is no more than 24 hours.
The maximum of $N_{\rm{swap}}$ can be determined by $N_{\rm{swap}}=T_{\rm{w}} \times 15 \times \cos\delta/2\psi$, where $\delta$ is the source declination, and $\psi$ is the detector's angular resolution.

%=================================================================
\section{Results on signal condition}\label{sec-4}
\begin{figure}
  \centering\includegraphics[width=0.7\linewidth]{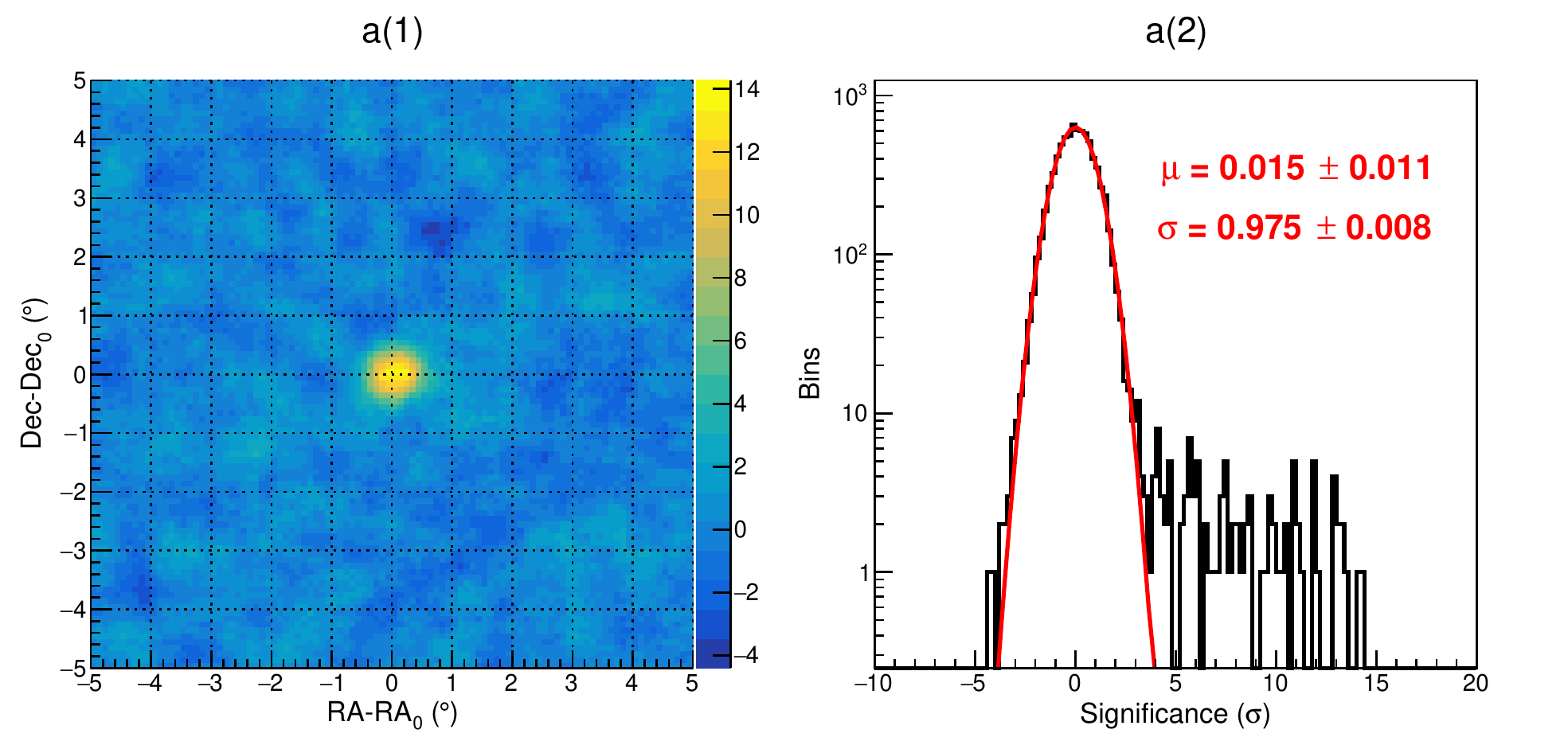}
  \centering\includegraphics[width=0.7\linewidth]{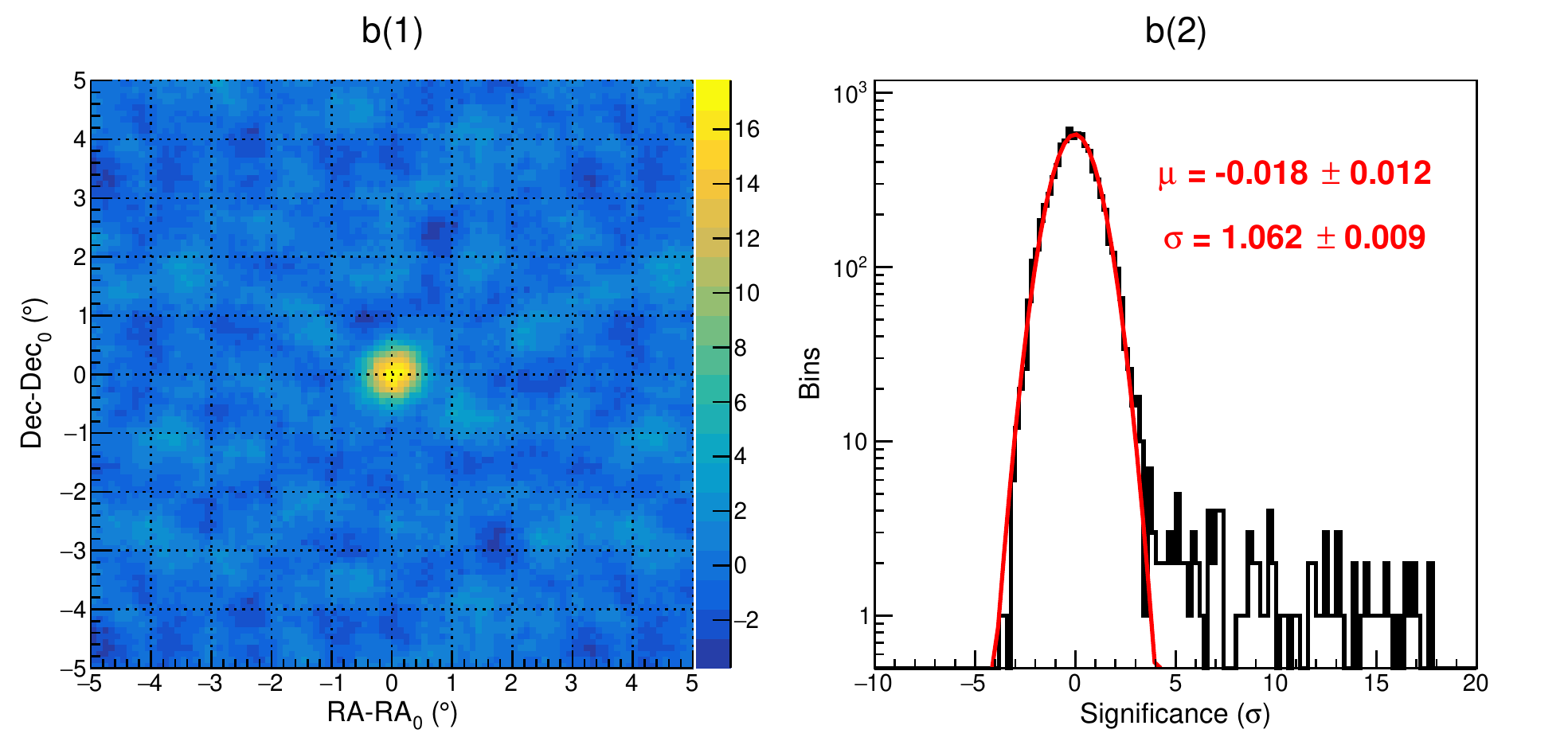}    
   \centering\includegraphics[width=0.7\linewidth]{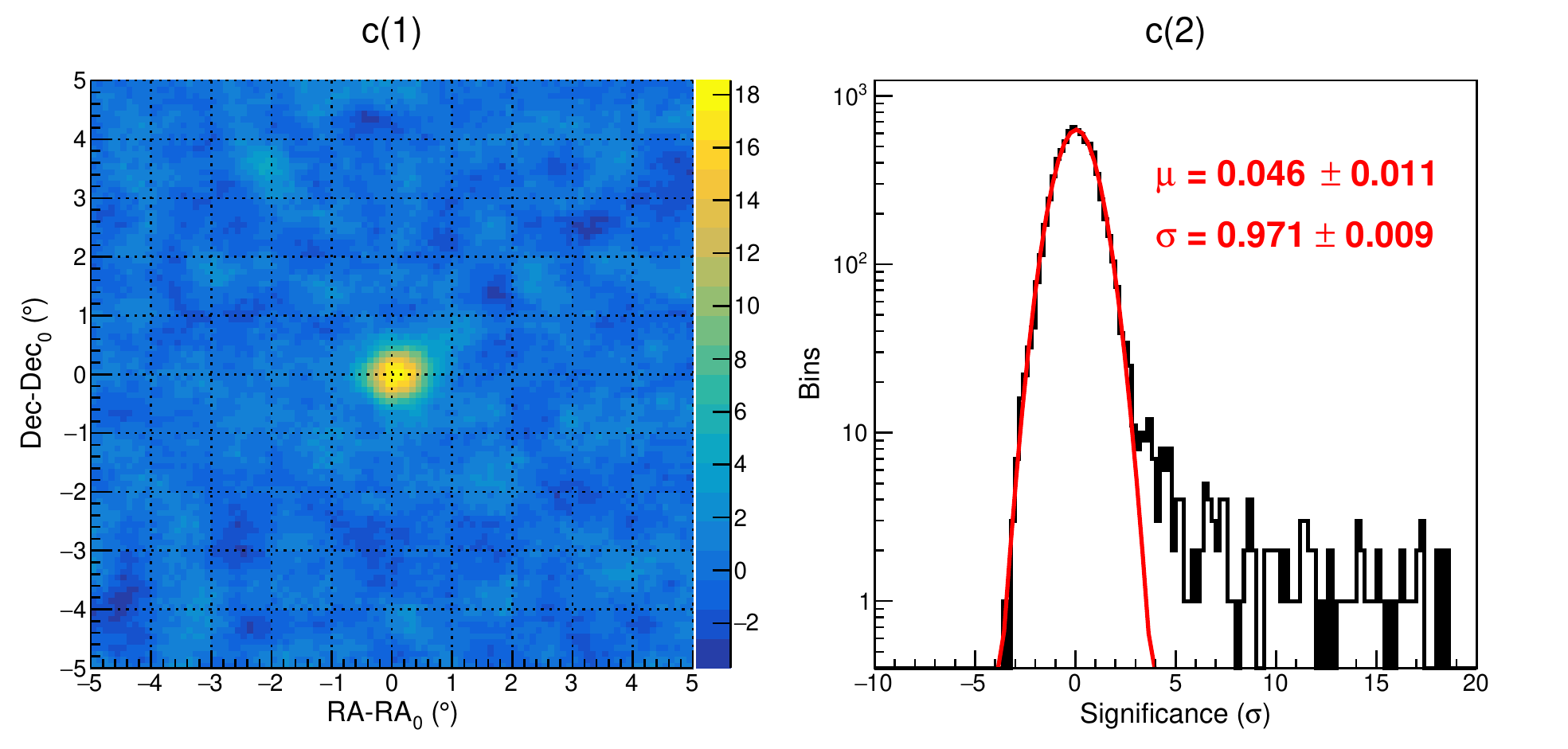}
  \centering\includegraphics[width=0.7\linewidth]{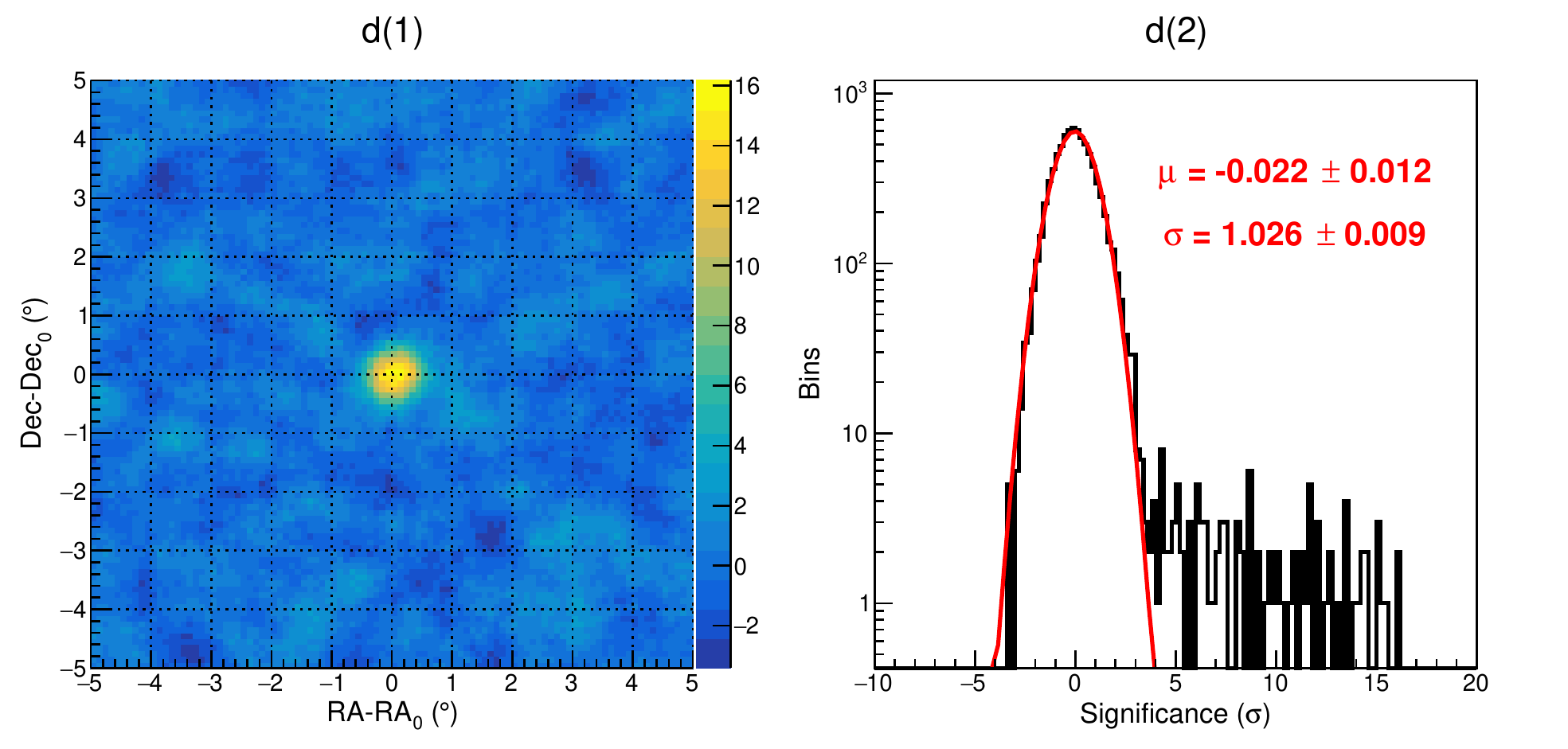}
  \caption{2-D significance sky maps and 1-D distributions of Crab significance using four background estimation methods in equatorial coordinates. 
  a(1)-d(1): maps of the statistical significance from equi-zenith angle, surrounding window, direct integration, and time-swapping method. The color scales on the right of the 2-D plots are the significance values.
    a(2)-d(2): 1-D distributions of significance from equi-zenith angle, surrounding window, direct integration, and time-swapping method. The red solid line represents the best Gaussian fit.}
\label{fig:Four_method1D2D}
\end{figure}
Based on generated simulation samples, $10^{\circ} \times 10^{\circ}$ sky maps with grid size of $0.1^{\circ} \times 0.1^{\circ}$ around the signal direction in the equatorial coordinate system are built for the different background methods, a $0.4^{\circ}$ smooth radius is used during the smoothing procedure, and the Li-Ma prescription~\cite{Li:1983fv} is used to calculate the significance of the signal: 
\begin{equation}
S =\sqrt{2}\{N_{\rm on}  \ln[\frac{1+\alpha}{\alpha}(\frac{N_{\rm on}}{N_{\rm on}+N_{\rm off}})]+N_{\rm off} \ln[(1+\alpha)(\frac{N_{\rm off}}{N_{\rm on}+N_{\rm off}})]\}^{1/2},
\end{equation}
where $N_{\rm on}$ is the number events in the on-source window, $N_{\rm off}$ is the number events in the off-source windows and $\alpha$ is the ratio of on-source time to off-source time. After the significance analysis, some interesting points can be drawn from the significance results as shown in Figures~\ref{fig:Four_method1D2D}-\ref{fig:Four_method}.

\begin{itemize}
\item Figure~\ref{fig:Four_method1D2D} shows the 2-D significance maps and 1-D distributions of significance by applying four different background estimation methods. From top to bottom, they are corresponding to equi-zenith angle, surrounding window, direct integration, and time-swapping methods, respectively. The maximum significances of the signal are 14.3, 17.8, 18.8, 16.2, and $\alpha$ values used in this work are 1/6, 1/60, 1/3500, and 1/6, respectively. The red line shows the Gaussian fitting result for the 2-D significance value in the left panel located outside of a $1^{\circ}$ circular region centered on the source position. with $\mu= [0.015, -0.018, 0.046, -0.022]$, $\sigma=[0.975, 1.062, 0.971, 1.026]$. 
One can clearly see that all 4 background methods have values of $\mu=0$ and $\sigma=1$ and they are consistent within the statistical uncertainties.
\begin{figure}
  \centering\includegraphics[width=0.32\linewidth]{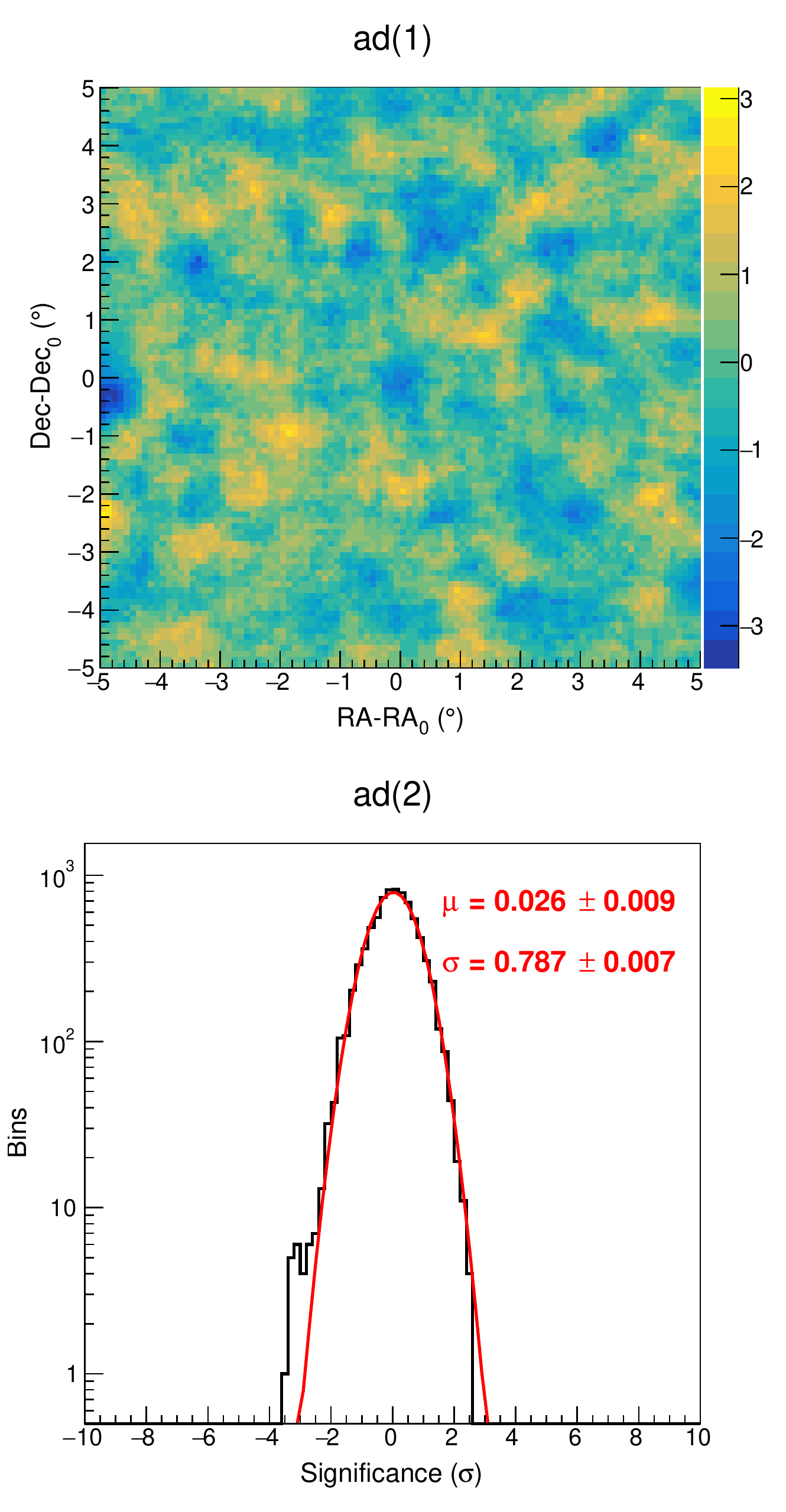}
  \centering\includegraphics[width=0.32\linewidth]{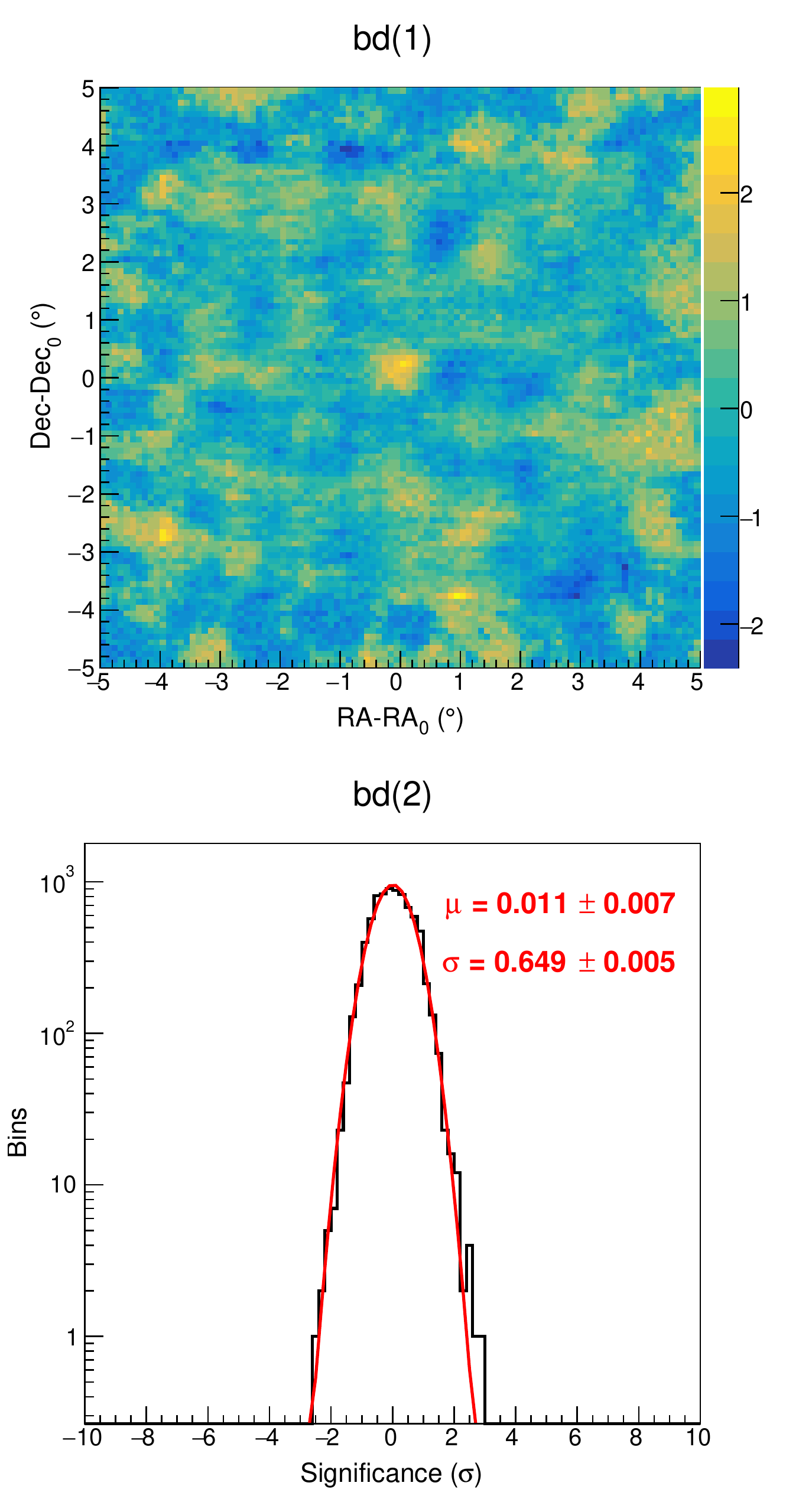}
  \centering\includegraphics[width=0.32\linewidth]{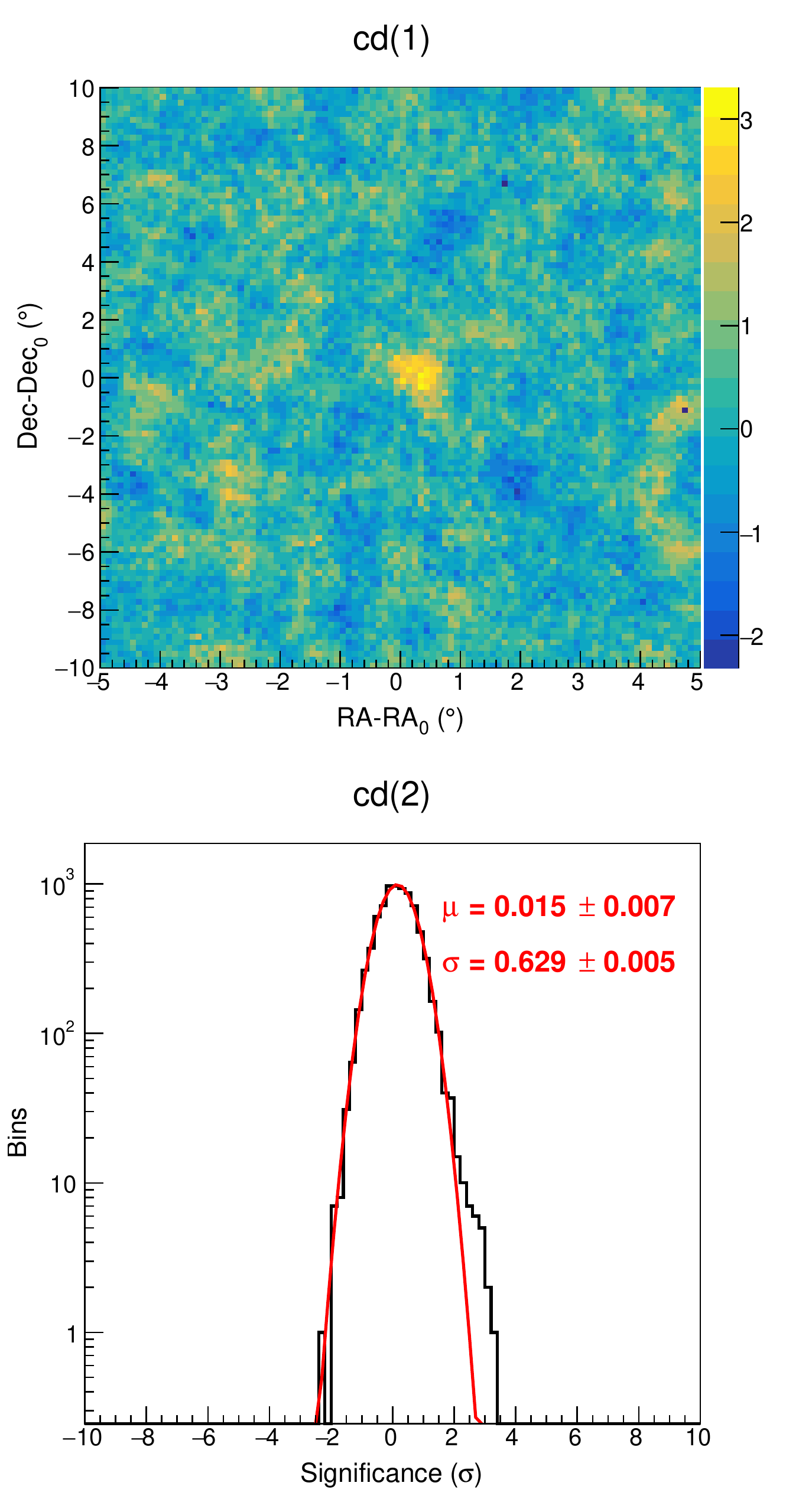}
  \caption{2-D and 1-D significance differences between different background estimation methods. The color scales on the right of the 2-D plots are the significance difference values, and the bottom three 1-D histograms are the distributions of all the points in the corresponding upper 2-D plots. Left panel: significance difference between equi-zenith angle method and time-swapping method. Middle panel: significance difference between surrounding window method and time-swapping method. Right panel: significance difference between direct integration method and time-swapping method.}
\label{fig:diff_sig}
\end{figure}

\item In order to further check the difference among methods, the result of the time-swapping method is taken as the reference method. Figure \ref{fig:diff_sig}~shows the 2-D (top panel) maps of the bin-by-bin subtraction from 2-D significance maps between the time-swapping method and the other three methods. The 1-D (bottom panel) maps in Figure~\ref{fig:diff_sig}~are the distributions of all the points in the corresponding upper 2-D 
significance difference plots. Their distributions can be well fitted with a Gaussian function with $0.01<\mu<0.02$, $0.63<\sigma<0.79$. 
We can conclude that no bias is introduced by any of the background estimation methods and that the systematic uncertainty is around $\sigma$=0.7.

\item Both equi-zenith angle and time-swapping methods use the same $\alpha$ value, 1/6, while the calculated maximum significance of the former method is 88\% that of the latter method, as shown in a(2) and d(2) of the Figure~\ref{fig:Four_method1D2D}.
This can be explained by an additional cut on the zenith angle due to avoiding off-source windows overlapping effect in the equi-zenith angle method.
 A simple calculation confirms about 14\% signal is missed during the analysis, this number is consistent with the significance difference.

\begin{figure}
\centering\includegraphics[width=0.7\linewidth]{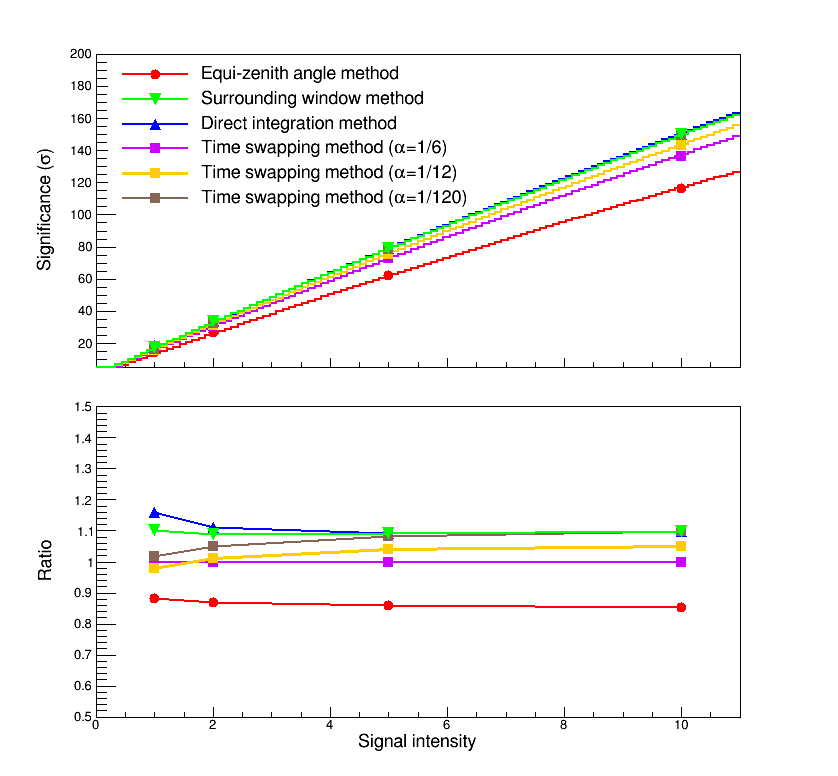}
\caption{Comparison of four background estimation methods with 1, 2, 5, and 10 times signal intensity for signal events from the Crab Nebula.
Top panel: significance of four background estimation methods. The red circles stand for the equi-zenith angle method, green inverted triangles denote the surrounding window method, and blue triangles show the direct integration method. The violet squares, orange squares and brown squares represent the time-swapping method when $\alpha=1/6$, $\alpha=1/12$, and $\alpha=1/120$, respectively. Bottom panel: significance ratio of above all results to that of the time-swapping method.}
\label{fig:Four_method}
\end{figure}
\item In order to further check this tendency relative to the signal intensity, we artificially increase the rate of Crab Nebula signal events, and a similar study of the significance calculation is carried out for increasing signal intensity for factors of 2, 5, and 10. The result is shown in Figure~\ref{fig:Four_method}. The top panel shows the calculated significance from different background estimation methods as a function of signal intensity, and the lines represent the value from Li-Ma formula. A non-linear increasing tendency is observed.

\item From the relationship between the maximum significances and $\alpha$ values of the four background estimation methods, one can easily find that a larger significance is obtained for decreasing values of $\alpha$. For further examination this point, 3 different $\alpha$ values are compared in the time-swapping method, which are 1/6, 1/12, and 1/120. The significance results are shown in Figure~\ref{fig:Four_method}. A larger significance is obtained for $\alpha=1/120$, which is quite reasonable. In principle, smaller $\alpha$ means that a bigger background region is used, and less fluctuation should be guaranteed with stable acceptance.
\end{itemize}

\section{Results on no signal condition}\label{sec-5}
\begin{figure}
  \centering\includegraphics[width=0.45\linewidth]{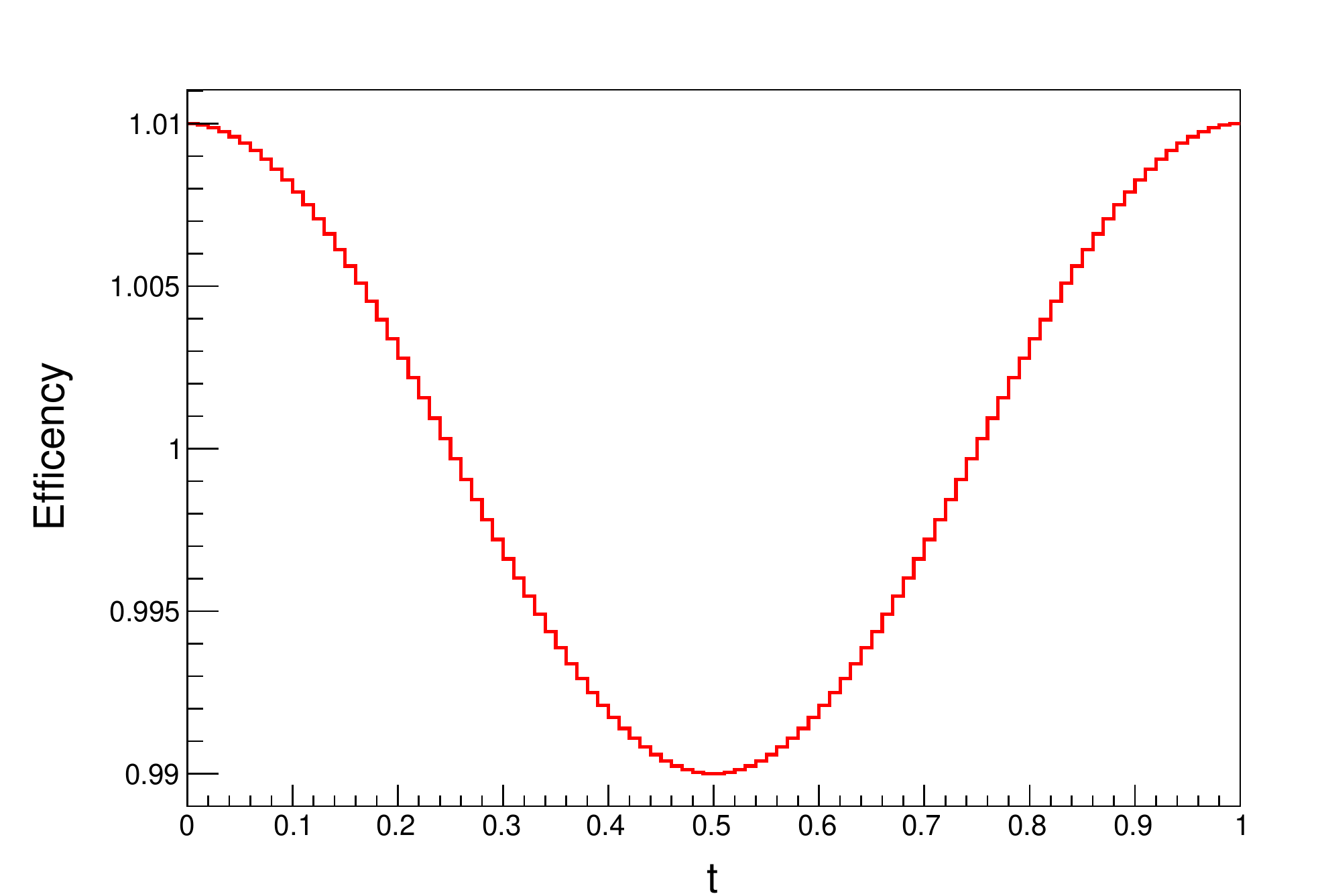}
    \centering\includegraphics[width=0.45\linewidth]{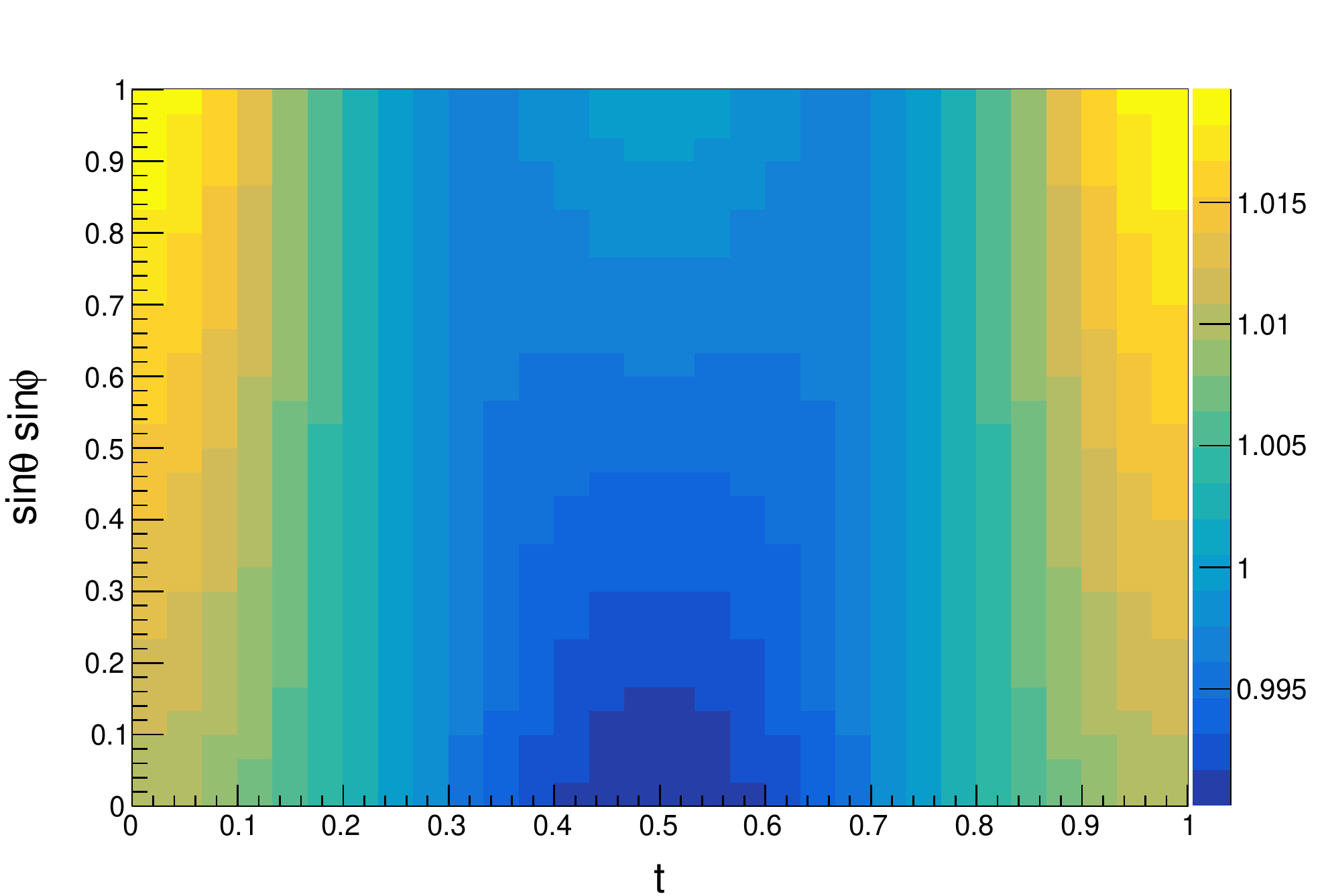}
  \caption{Efficiency maps of the acceptance variation under different condition.
  Left panel: the acceptance of the detector variation with time overall.
  Right panel: the acceptance of the detector variation with both space and time, the color scale represents the efficiency of the acceptance variation.}
  \label{fig:an_bn}
\end{figure}

In this section, we present details about the influence due to different background estimation methods and their applicability in the case of no signal. 
Sometimes, the EAS detector is unstable. For example in paper~\cite{ARGO-YBJ:2009tho}, one can see a variation of detector efficiency with time from ARGO-YBJ experiment, which is well described by the sine/cosine formula, and it is easy to keep such variation as a kind of daily modulation. 
Following this, two additional detector instability assumptions are introduced in this work:
\begin{enumerate}
\item  The detector acceptance varies with time. i.e. $R_{1}(\theta, \phi, t) = \mathrm{A_{1}}(\theta, \phi,t){\rm{a}_{n}}$.
\item  The detector acceptance varies both with time and in the direction. i.e. $R_{2}(\theta, \phi, t) = \mathrm{A_{2}}(\theta, \phi, t)\rm{b}_{n}$.
\end{enumerate}
\begin{table}[H]
\caption{Functions of the acceptance variation in different cases. The function $\rm{a_{n}}$ denotes the acceptance variation over time overall. The function $\rm{b_{n}}$ describes the acceptance variation with both space and time.}
\begin{center}
\renewcommand\arraystretch{1.2}
\begin{tabular}{ccccccc}
\hline
\hline
Function    &  Expression                    & n &  $\Lambda$   \\
\hline
 $\rm{a_{n}}$ & $1+ \mathrm{\Lambda}\cos(2\pi t)$        & 0 & 0  \\ 
             &                                            & 1 & 1\% \\ 
             &                                            & 2 & 2\% \\
             &                                            & 3 & 5\% \\
             &                                            & 4 & 10\%   \\     
\hline            
$\rm{b_{n}}$ & $1+\mathrm{\Lambda}\cos(2\pi t)[1+\sin \theta \sin \phi \cos(2\pi t)]$  & 1 & 1\% \\                           
             &                                                                               & 2 & 2\% \\ 
             &                                                                               & 3 & 5\% \\ 
             &                                                             	 	  & 4 & 10\%   \\ 
             
\hline
\hline
\end{tabular}
\end{center}
\label{table:function}
\end{table}
The acceptance variation functions of the detector in this work are constructed as shown in Table~\ref{table:function}, where $\Lambda$ is the acceptance change ratio from 1\% to 10\%, $t$ is the time during a sidereal day from 0 to 1, $\theta$ is the zenith angle of the samples from $0^{\circ}$ to $90^{\circ}$, and $\phi$ is the azimuth angle generated by uniform sampling in the interval $0^{\circ}-360^{\circ}$.
The acceptance variation maps of the detector for the two assumptions with $\Lambda=1\%$ are shown in Figure~\ref{fig:an_bn}. 
 
Based on the simulation samples, 1-D significance distributions have been built using the different background estimation methods. 
For the no signal condition, the 1-D significance distribution should be described by a standard normal distribution ($\mu=0$, $\sigma=1$, $\chi^{2}/\rm{dof}=1$). Thus in the following discussion, $\chi^{2}/\rm{dof}$ is an indicator of the goodness of fit. The values of $\chi^{2}/\rm{dof}$ obtained by using the standard normal distribution to fit the corresponding 1-D significance distribution from different background estimation methods are listed in Tables~\ref{table:chi-ndf_total}-\ref{table:chi-ndf_swap}. 
\begin{table}[H]
\begin{center}
\caption{Under different acceptance variation functions, the $\chi^{2}$/dof of the 1-D significance distribution from four background estimation methods relative to the standard normal distribution.}
\label{table:chi-ndf_total}
\begin{tabular}{rccccccc}
\hline
\hline
&Function &  Equi-zenith angle & Surrounding window & Direct integration & Time-swapping\\
&               &                               &                                  &($T_{\rm{w}}$=24 hours)&($T_{\rm{w}}$=24 hours)\\
\hline
& $\mathrm{a}_{0}$ &1.29&0.69&1.31&1.56\\   
& $\mathrm{a}_{1}$ &2.70&0.86&1.31&1.36\\   
&$\mathrm{a}_{2}$ & 2.11&0.76&1.45&1.24\\
&$\mathrm{a}_{3}$ &2.84&0.78&1.44&1.11\\
&$\mathrm{a}_{4}$ &3.74&0.94&1.82&24.29\\
\hline 
&$\mathrm{b}_{1}$ &1.93&1.13&10.58&3.57\\
&$\mathrm{b}_{2}$ &4.28&0.82&92.77&61.70\\
&$\mathrm{b}_{3}$ &16.86&0.86&1578.99&1091.93\\
&$\mathrm{b}_{4}$ &71.64&0.97&7530.80&5755.99\\
\hline
\hline
\end{tabular}
\end{center}
\end{table}

\subsection{$\chi^{2}/\rm{dof}$ evaluation in case 1 assumption $R_{1}(\theta, \phi, t)$ } 
In this assumption, for a variation of acceptance $\Lambda$ up to 5\%, no obvious difference exists among the four methods with $\chi^{2}/\rm{dof}$ in the range of 0.78-3.47. In other words a variation in the detector acceptance at the level of 5\% can be handled by the four methods. More details are given in Table~\ref{table:chi-ndf_total}.
\begin{figure}
  \centering\includegraphics[width=0.8\linewidth]{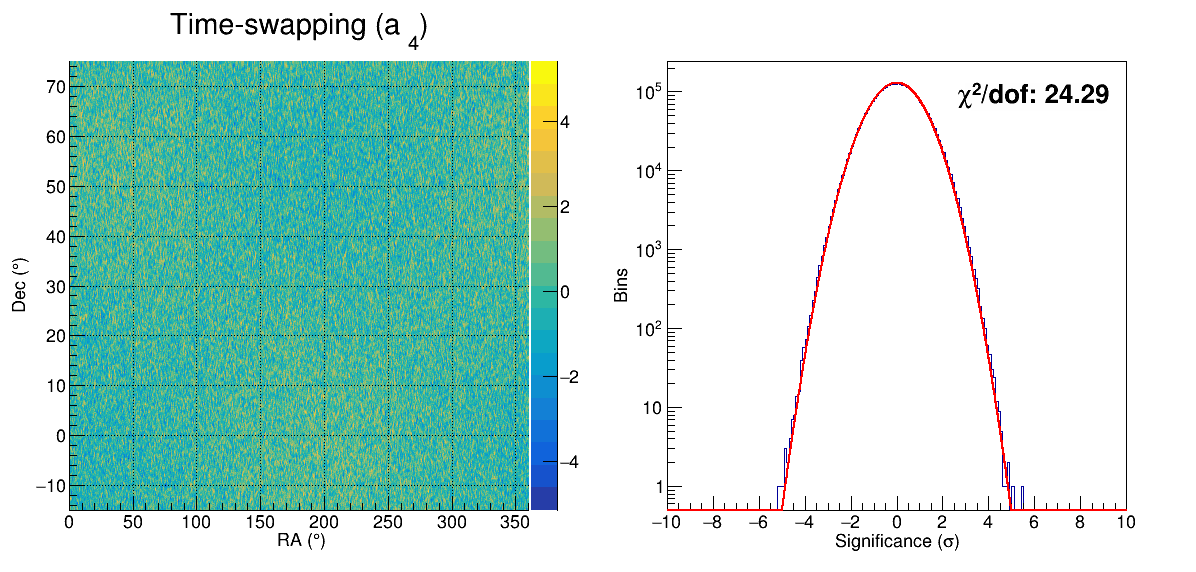}
  \caption{2-D significance sky maps and 1-D significance distributions from the time-swapping method under $\rm{a}_{4}$ assumption.}
  \label{fig:Time-swapping_a4}
\end{figure}
For the equi-zenith angle method, a little larger $\chi^{2}/\rm{dof}$ value from 2.70 to 3.74 is obtained when the acceptance changes from 1\% to 5\%. One possible reason is due to its limited statistics, here only the $10^{\circ} \times 10^{\circ}$ sky map is used in the analysis. 

The $\chi^{2}/\rm{dof}$ of the time-swapping method is 24.29 when the acceptance changes $10\%$, and the 2-D and 1-D significance maps are shown in Figure~\ref{fig:Time-swapping_a4}. 
From the 2-D distribution, a few clustering blocks can be seen. Such as in $150^{\circ}$<RA <$200^{\circ}$ and Dec<$20^{\circ}$, the significance is high, and that is low in $150^{\circ}$<RA <$200^{\circ}$ and Dec>$20^{\circ}$. It basically means this background estimation is
not good enough, thus a large $\chi^{2}/\rm{dof}$ value of 24.29 is obtained. At this time, it violates the principle of this method, and we will continue to study this point in our future work.

\subsection{$\chi^{2}/\rm{dof}$ evaluation in case 2 assumption $R_{2}(\theta, \phi, t)$}
Table~\ref{table:chi-ndf_total} shows that the $\chi^{2}/\rm{dof}$ of the surrounding window method is rather stable around 1 in this assumption, which is natural for this method. One advantage of this method is that the effect due to time variation and direction can be minimized. In this method, a hollow rectangular region surrounding the center cell of interest is used for background estimation. Pollution from events in the center cell of interest is avoided, and the acceptance ratio hardly depends on the acceptance changes of the center cell of interest. 
\begin{figure}
  \centering\includegraphics[width=0.8\linewidth]{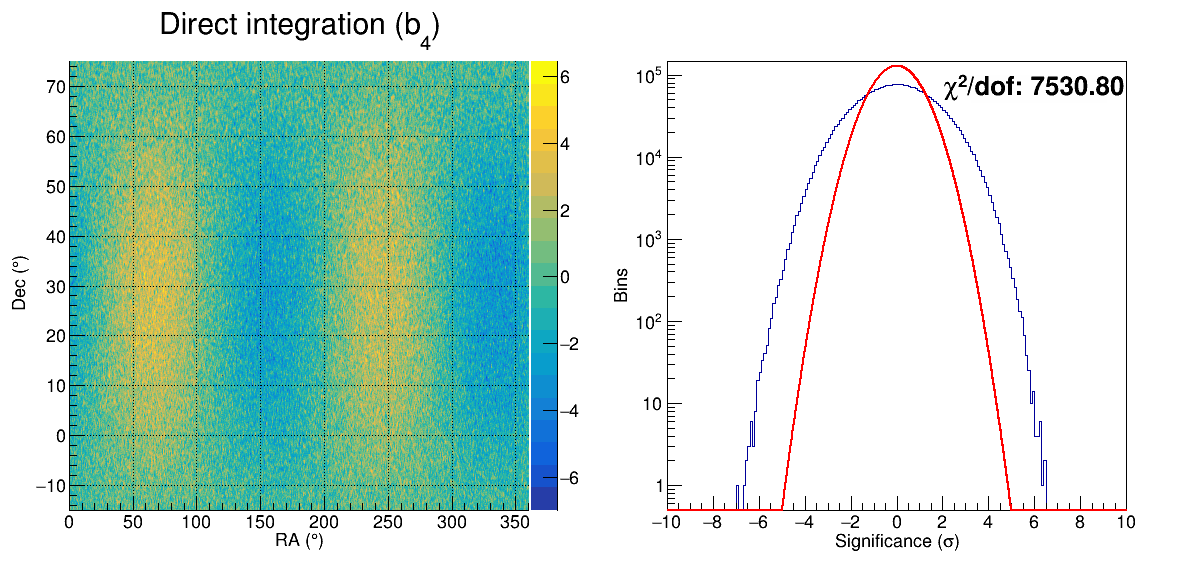}
  \caption{2-D significance sky maps and 1-D significance distributions from the direct integration method under $\rm{b}_{4}$ assumption.}
  \label{fig:b4}
\end{figure}

For the equi-zenith angle method, when the acceptance changes from 5\% to 10\%, the $\chi^{2}/\rm{dof}$ changes from 1.93 to 71.64. The main reason is that the lack of an azimuth correction leads to the ineffectiveness of the equi-zenith angle method. In this case, the azimuth angle distribution varies with time every second, and it is impossible to make an azimuth correction at every moment.

For the direct integration and time-swapping methods, under the time windows $T_{\rm{w}}$ of 24 hours, the $\chi^{2}/\rm{dof}$ of them gets worse when the acceptance changes $2\%$. When the variation of acceptance $\Lambda$ up to 10\%, the $\chi^{2}/\rm{dof}$ of the direct integration method can even reach 7530.8, and the 2-D and 1-D significance maps are shown in Figure~\ref{fig:b4}. Here the detector acceptance is not only local-direction dependent, but also shows strong time-dependent variation, which simply violates the basic idea of the two methods.

\subsection{Time window effect in case 2 assumption}
As discussed above, the $\chi^{2}/\rm{dof}$ of the direct integration and time-swapping methods get even worse when the acceptance changes significantly with both space and time. In this acceptance assumption, in order to further study whether the two methods can be improved with different time windows, the background estimation of them under $T_{\rm{w}}=$1 hour, 2 hours, 4 hours, and 8 hours have also been studied.

\begin{table}[H]
\begin{center}
\caption{Under different time windows and different acceptance variation functions, the $\chi^{2}/\rm{dof}$ of the 1-D significance distribution from the direct integration method relative to the standard normal distribution.}
\label{table:chi-ndf_direct}
\begin{tabular}{rccccccc}
\hline
\hline
   Function& 1 hour & 2 hours& 4 hours& 8 hours& 24 hours\\
   \hline    
 $\mathrm{b}_{1}$ &5020.45&10.78&1.36&4.34&10.58\\
 $\mathrm{b}_{2}$ &5124.03&10.82&1.98&45.03&92.77\\ 
 $\mathrm{b}_{3}$ &4953.92&8.36&82.75&900.06&1578.99\\ 
 $\mathrm{b}_{4}$ &4778.79&2.89&809.41&4679.31&7530.80\\
\hline
\hline
\end{tabular}
\end{center}
\end{table}
From the results listed in Table~\ref{table:chi-ndf_direct}, when the acceptance changes 2\% (corresponding to the case $\rm{b_{2}}$), with the time window of the direct integration method shrinks, $\chi^{2}/\rm{dof}$ turns smaller firstly from 92.77 at 24 hours to 1.98 at 4 hours, then turns larger to 5124 at 1 hour. One possible explanation is for a short time window, such as 1 hour, the on-source window and off-source windows are closely correlated and a poor fit is obtained. In this way, one can see that the time window of 1 hour (covering $15^{\circ}$ in HA) and 4 hours (covering $60^{\circ}$ in HA) are quite different for the background estimation. Comparing with the 1-hour background, the 4-hour background covers a wider range, thus the on-source window and off-source windows have less correlation.
For time windows more than 4 hours, the larger the time window, the larger the influence of the background estimation has, and the $\chi^{2}/\rm{dof}$ value is bad. For the time window of 4 hours, the background estimation statistic is reduced to achieve some balance with the acceptance change, which can reduce the impact on the background estimation to some extent, and a relatively minimal $\chi^{2}/\rm{dof}$value of 1.98 is obtained.

\begin{table}[H]
\begin{center}
\caption{Under different time windows and different acceptance variation functions, the $\chi^{2}$/dof of the 1-D significance distribution from the time-swapping method relative to the standard normal distribution.}
\label{table:chi-ndf_swap}
\begin{tabular}{rccccccc}
\hline
\hline
Function& 1 hour & 2 hours& 4 hours& 8 hours& 24 hours\\
\hline
 $\mathrm{b}_{1}$ &1.60&1.44&1.34&3.84&3.57\\
 $\mathrm{b}_{2}$ &1.55&1.09&2.11&23.50&61.70\\
 $\mathrm{b}_{3}$ &4.47&3.76&89.08&696.80&1091.93\\
 $\mathrm{b}_{4}$ &45.53&64.45&808.65&3883.58&5755.99\\
 \hline
 \hline
\end{tabular}
\end{center}
\end{table} 

Results of the time-swapping method in different time windows is displayed in Table~\ref{table:chi-ndf_swap}.
It shows that when the time window is taken from 1 hour to 24 hours in the case of 2\% change in acceptance, the value of $\chi^{2}/\rm{dof}$ changes from 1.55 to 61.70. As is the case for the direct integration method, the time-swapping method is also affected by the time window.  
We find that for variations in the acceptance range up to 2\%, the time-swapping method is sensitive to the selection of time window, and that the time window of 1 hour is less affected for the background estimation. 
In this case, a 2-D distribution of all events in hour angle coordinate (HA, Dec) is accumulated for every 1 hour. For each event, a "fake background event" is generated by calculating a new value of (RA, Dec) through associating the event time with a new direction of (HA, Dec) chosen randomly from above 2-D distribution, and this procedure is repeated 6 times. In this method, the short time window naturally compensates for the acceptance variation.
However, when the acceptance varies significantly over time and space, the basic principle of the method to estimate the background is violated even with the 1-hour time window. For example, for an acceptance change higher than 5\%, we find that shortening the time window has no effect on the large $\chi^{2}/\rm{dof}$ value.

\section{Summary and discussion}\label{sec-6}
When using a ground array to detect TeV $\gamma$-ray signals, it is necessary to estimate the background accurately. 
In this paper, a detailed comparison study of four different background estimation methods, equal-zenith method, surrounding window method, direct integration method, and time-swapping method are made using simulation samples. Although the detailed implementations are different, with the variation of parameters and corresponding corrections, it is interesting to see an agreement between the methods at the 1$\sigma$ level in the case of a signal condition. Moreover, no bias is observed in the background estimation. From the calculated significance point of view, the smaller $\alpha$ (the ratio of on-source time to off-source time) parameter, the higher significance value obtained, and the rate of increasing significance is not linearly dependent on the signal intensity.

In the case of the no signal condition, when the acceptance of the detector only changes with time and this variation is less than 5\%, no bias is observed among the four background estimation methods. However, when the acceptance of the detector changes with both space and time, except for the surrounding window method, the other three methods are not applicable. Additionally, a study of the effect of the time window on the direct integration and time-swapping methods is made. We find that the background estimations in the the two methods are sensitive to the selection of time window. Compared with the time windows of 1 hour, 2 hours, 8 hours, and 24 hours, the time window of 4 hours in the direct integration method and that of 1 hour in the time-swapping method are suitable options, which can reduce the impact on the background estimation to some extent. However, when the acceptance of the detector changes significantly with both space and time, no improvement in the background estimation can be obtained even if the time window is short.
We hope that above detailed analysis can help to improve the understanding of the method application condition.

%=================================================================
\acknowledgments
We would like to thank Zhi-Guo Yao, Guang-Man Xiang, Liang Chen, and Yu-Hua Yao for helpful discussions.
This work was supported by the National Natural Fund Astronomical Joint Fund Project (Grant No. U1831208), NSFC (Grants Nos. 11675187,11375224, 11975072, 11835009, 11875102, and 11690021), the Joint Large-Scale Scientific Facility Funds of the NSFC, CAS (Grants Nos.U1332201 and U1532258), the Liaoning Revitalization Talents Program (Grant No. XLYC1905011), the Fundamental Research Funds for the Central Universities (Grant No. N2005030), the National 111 Project of China (Grant No. B16009), and the Science Research Grants from the China Manned Space Project (Grant No. CMS-CSST-2021-B01).

%\paragraph{Note added.}

%\bibliography{mybibfile}

\end{document}